\documentclass[prd,epsf,amsmath,superscriptaddress,citeautoscript,showpacs,floatfix,twocolumn,nofootinbib,preprintnumbers]{revtex4-2}
\usepackage[T1]{fontenc}
\usepackage{txfonts}

\usepackage[dvipsnames]{xcolor}
\definecolor{red}{rgb}{0.9, 0,0}
\definecolor{cerulean}{rgb}{0., 0.42,0.9}
\definecolor{navy}{rgb}{0.05, 0.05,0.8}

\usepackage[colorlinks]{hyperref}
\hypersetup{
    colorlinks = true,
    citecolor  = navy,
	linkcolor  = navy
}

\usepackage{slashed}
\usepackage{graphics}
\usepackage{amsmath}
\usepackage{amssymb}
\usepackage{latexsym}
\usepackage{mathtools}
\usepackage{dsfont}
\usepackage{cleveref}
\usepackage{hyperref}
\usepackage{amsfonts}
\usepackage{enumitem}
\usepackage{xstring} 
\usepackage{xspace} 
\usepackage[normalem]{ulem}
\usepackage{fontawesome}
\usepackage{braket}
\usepackage{mathrsfs}

\newcommand{\comment}[1]{}

\usepackage{subfiles} 

\begin{document}

\title{Decoherence by warm horizons}

\author{Jordan Wilson-Gerow}
    \affiliation{Walter Burke Institute for Theoretical Physics, California Institute of Technology, Pasadena, CA 91125, USA}
    \affiliation{Theoretical Astrophysics 350-17, 
		California Institute of Technology, Pasadena CA 91125, USA}	
    \author{
	Annika Dugad}
    \affiliation{Walter Burke Institute for Theoretical Physics, California Institute of Technology, Pasadena, CA 91125, USA}
    \affiliation{Theoretical Astrophysics 350-17, 
		California Institute of Technology, Pasadena CA 91125, USA}	
	\author{
	Yanbei Chen}
    \affiliation{Walter Burke Institute for Theoretical Physics, California Institute of Technology, Pasadena, CA 91125, USA}
    \affiliation{Theoretical Astrophysics 350-17, 
		California Institute of Technology, Pasadena CA 91125, USA}

\date{May 1, 2024}
	
	\begin{abstract}
		
Recently Danielson, Satishchandran, and Wald (DSW) have shown that quantum superpositions held outside of Killing horizons will decohere at a steady rate. This occurs because of the inevitable radiation of soft photons (gravitons), which imprint a electromagnetic (gravitational) ``which-path'' memory onto the horizon. Rather than appealing to this global description, an experimenter ought to also have a local description for the cause of decoherence. One might intuitively guess that this is just the bombardment of Hawking/Unruh radiation on the system, however simple calculations challenge this idea---the same superposition held in a finite temperature inertial laboratory does not decohere at the DSW rate. In this work we provide a local description of the decoherence by mapping the DSW set-up onto a worldline-localized model resembling an Unruh-DeWitt particle detector. We present an interpretation in terms of random local forces which do not sufficiently self-average over long times. Using the Rindler horizon as a concrete example we clarify the crucial role of temperature, and show that the Unruh effect is the only quantum mechanical effect underlying these random forces. A general lesson is that for an environment which induces Ohmic friction on the central system (as one gets from the classical Abraham-Lorentz-Dirac force, in an accelerating frame) the fluctuation-dissipation theorem implies that when this environment is at finite temperature it will cause steady decoherence on the central system. Our results agree with DSW and provide the complementary local perspective.

	\end{abstract}
	\preprint{CALT-TH 2024-017}

\maketitle

%

\section{Introduction}
\label{sec:intro}

\subsection{Motivation}

Through a beautiful stroke of insight, Danielson, Satishchandran, and Wald (DSW) arrived at the conclusion that quantum superpositions held near Killing horizons will suffer a steady rate of decoherence~\cite{Danielson:2022tdw,Danielson:2022sga}. Their intuition came from fundamental considerations regarding the consistency of causality with the fact that long-ranged Newton and Coulomb fields can entangle particles. Their logical conclusion: if causality is to be preserved while Alice and Bob attempt to perform EPR-like tests on either side of a horizon, there must be a pervasive source of decoherence conspiring to disrupt the experiments~\cite{Danielson:2021egj,Gralla:2023oya}.

The effect is exemplified by a lab which accelerates with proper acceleration $a$, containing a particle of charge $q$, which is held in a superposition of two states separated by distance $\varepsilon$. With respect to the lab's proper time, the system decoheres with a constant rate\footnote{For brevity, here and throughout, we refer to the decoherence rate  as $\Gamma^{\rm EM}_{\rm DSW}$, however it should be understood that DSW first determined the scaling whereas Gralla and Wei~\cite{Gralla:2023oya} first computed the precise numerical coefficients.}
\begin{equation}\label{eq:DSWEMrate}
    \Gamma^{\rm EM}_{\rm DSW}=\frac{q^{2}a^{3}\varepsilon^{2}}{12\pi^{2}}\,.
\end{equation}
That is, the visibility of interference fringes is suppressed by $\exp(-T\Gamma^{\rm EM}_{\rm DSW})$, where $T$ is the total elapsed proper time.

DSW provided a slick argument for the decoherence by demonstrating the existence of electromagnetic and gravitational memory effects for Killing horizons, wherein certain processes involving charged (massive) particles would imprint a net DC change in the electromagnetic (gravitational) wave profile on the horizon.   As it is a DC shift, it is explained quantum mechanically by the radiation of \emph{soft} photons and gravitons through the horizon. 

From a global perspective the decoherence is intuitive---the two branches of the superposition radiate soft photons carrying ``which-path'' information through the horizon. It remains unclear, however, how the experimenter would explain the effect in their local frame. A natural first guess is that this is due to Unruh/Hawking radiation, however simple estimates suggest that thermal photons cannot be the cause. If the decoherence were to be explained by the scattering of thermal photons, then one expects that the rate would depend on the Thompson cross section. As already noted by DSW, this cross-section scales inversely with the mass of the charged particle and can be made arbitrarily small relative to Eq.~\eqref{eq:DSWEMrate}.

This however raises a puzzle: how could the experimenter in the accelerating lab \emph{locally} explain this persistent decoherence if it isn't coming from the Unruh effect?  In the language of open quantum systems, rather than describing the decoherence as arising from the central system radiating zero-energy modes into the bath, we ought to be able to explain it locally as a physical action of the bath onto the central system---but what is this action if not simply thermal fluctuations?

In this paper we aim to address this question and provide an intuitive explanation for the steady rate of decoherence from the perspective of the accelerated observer. Despite the above hesitation, we demonstrate that from their perspective it can in fact be explained by well-known properties of classical electromagnetism plus the existence of a finite temperature due to the Unruh effect. Our results agree precisely with DSW, providing a complementary perspective on the effect. Furthermore, the essential role played by temperature corroborates the findings of Gralla and Wei~\cite{Gralla:2023oya}, who demonstrated the vanishing of the DSW decoherence for extremal black holes.  We primarily restrict to the case of electromagnetic radiation and Rindler horizons, however this is sufficient to illustrate the general lesson.

\subsection{Equivalence between DSW and Unruh-Dewitt}

The main statement in its heuristic form is: if coupling a system to a certain environment induces Ohmic friction (a force term given by a first time derivative) in the system's classical equation of motion, then when that environment is at finite temperature it will have a spectrum of thermal fluctuations that cause a constant rate of decoherence in the system. From this, and the fact that in an accelerated frame the relativistic Abraham-Lorentz-Dirac (ALD) radiation reaction force contains an Ohmic piece, it follows that a finite Unruh temperature will lead to decoherence.

The physical picture turns out to be somewhat different than a typical collisional decoherence model; rather, it is dephasing which dominates. The experimenter prepares a superposition of a particle with charge $q$ delocalized in position space, $|\psi\rangle=|+\varepsilon/2\rangle+|-\varepsilon/2\rangle$. In the absence of electric fields this state can evolve without acquiring a relative phase between branches. However, in the presence of an electric field fluctuation such that $q\, \delta E>0$, the state $|-\varepsilon/2\rangle$ sits at a slightly higher energy $\delta H\approx q\varepsilon\,\delta E$ than the state $|+\varepsilon/2\rangle$. Provided that the electric field fluctuation is sufficiently coherent over the duration of the experiment, the two states accumulate a relative phase due to their relative energy difference. While coherent over a single experiment, the long-lived mode $\delta \vec{E}$ is still a random variable and will be completely randomized over many such experiments, leading to a measured loss of interference contrast. This dephasing between the two branches of the wavefunction does not depend on the mass of the charged particle.

Ohmic baths are particularly relevant here, because the fluctuation-dissipation theorem ensures that they have the requisite long-time correlated fluctuations. Electric fields near a Rindler horizon, as we will demonstrate, behave as Ohmic baths. This then leads to the interesting observation, that for a single-shot interference measurement near a horizon it will appear as if the vacuum has spontaneously selected a preferred nonzero static electric field about which it fluctuates.

Before proceeding to the detailed computations, it is interesting to try to provide an order-of-magnitude estimate of the DSW decoherence rate using elementary physics - although there is a ``pitfall'' in this naive process. For two paths separated by a displacement vector {\boldmath{$\varepsilon$}}, their phase difference in an EM field is given by 
\begin{equation}
    \Delta \Phi =q\int d\tau  \hat{\mathbf{E}}(\tau) \cdot  \mbox{\boldmath $\varepsilon$}(\tau) \,,
\end{equation}
and the loss in contrast is given by 
\begin{equation}
\langle \Delta\Phi^2\rangle =q^2 \iint  d\tau_1 d\tau_2 \langle \hat E_I(\tau_1) \hat  E_J(\tau_2)\rangle \varepsilon^I(\tau_1)\varepsilon^J(\tau_2)\,.
\end{equation}
For steady state, we can write 
\begin{equation}
\label{generalW}
\langle \Delta\Phi^2\rangle =q^{2}\int \frac{d\Omega}{2\pi}
S_E^{IJ}(\Omega) \tilde\varepsilon_I(\Omega)\tilde\varepsilon^*_J(\Omega),
\end{equation} 
where $S_E^{IJ}$ is the {\it symmetrized} spectral density: 
\begin{equation}
\langle 
\tilde E^*_I  (\Omega)
\tilde E_J (\Omega')\rangle_{\rm sym} = 2\pi\delta(\Omega-\Omega') S_E(\Omega).
\end{equation}

Suppose $\varepsilon$ remains constant for a long duration $T$.  As a result, $\tilde \varepsilon$, the Fourier transform of $\varepsilon$, is highly bandlimited near $\Omega\sim 0$. We can convert Eq.~\eqref{generalW} into
\begin{equation}
\langle \Delta\Phi^2\rangle \approx q^{2}S_E^{IJ}(0)  \int \frac{d\Omega}{2\pi}\tilde\varepsilon^*_I(\Omega)
\tilde\varepsilon^*_J(\Omega)=  q^{2}\varepsilon_I \varepsilon_J S_E^{IJ}(0) T\,,
\end{equation}
which corresponds to a decoherence rate
\begin{equation}
\Gamma  \propto  \varepsilon_I \varepsilon_J S_E^{IJ}(0)\,.
\end{equation}
This is very promising since it seems every steady bath leads to a constant decoherence rate.   The pitfall comes when we apply the Planck's Law for thermal radiation, which gives
\begin{equation}\label{eq:Wplanck}
    [S_E^{IJ}(\Omega)]_{\rm Planck}=\delta^{IJ}\frac{\Omega ^3}{6 \pi }\left(\frac{1}{2}+\frac{1}{e^{\beta\Omega}-1}\right)\,,
\end{equation}
where $\beta =\hbar/(k_B T)$ is the inverse temperature.  This does not recover decoherence at a constant rate since $[S_E^{IJ}(0)]_{\rm Planck} = 0$.

This should be compared with the decoherence for an Unruh-DeWitt particle detector~\cite{Unruh:1976db, Hawking:1979ig} coupled to a finite temperature scalar field, for which the decoherence rate has the similar form
\begin{equation}
    \Gamma\propto S_{\phi}(0)=\lim_{\Omega\rightarrow 0} \frac{1}{2}\int d\tau\, e^{i\Omega\tau}\,\langle\{\phi(\tau),\phi(0)\}\rangle\,,    
\end{equation}
with
\begin{equation}
    S_{\phi}(\Omega)=\frac{\Omega}{\pi}\left(\frac{1}{2}+\frac{1}{e^{\beta\Omega}-1}\right)\,.
\end{equation}
This evidently has $S_{\phi}(0)\neq0$, and the particle detector suffers constant decoherence in a thermal bath.

Clearly, whether a thermal bath will or will not cause steady decoherence is determined by presence or absence of a term linear in frequency multiplying the Bose factor. As it turn out, this depends on the nature of a system's coupling to the environmental field, and crucially for this problem, it depends on the system's acceleration. As we will show, the spectrum of electric field fluctuations felt by the dipole in a uniformly accelerated frame is actually 
    \begin{equation}\label{eq:acceleratedpowerspectrum}
    [S_E^{IJ}(\Omega)]_{\rm accel.}=\delta^{IJ}\,\frac{a^{2}\Omega+\Omega ^3}{6 \pi }\left(\frac{1}{2}+\frac{1}{e^{\beta\Omega}-1}\right)\,,
\end{equation}
with $\beta=2\pi/a$ being the inverse Unruh temperature.\footnote{See also \cite{Boyer:1980wu, Boyer:1984yqq}, for analogous results, outside of the context of decoherence.} This leads to non-zero thermal fluctuations at arbitrarily low frequencies and thus a steady decoherence rate
\begin{equation}\label{eq:approxdecohrate}
    \Gamma\propto \frac{q^{2}|\varepsilon|^{2}a^{2}}{\beta}\,, 
\end{equation}
We will show that upon careful computation this line of reasoning matches the DSW decoherence rate exactly.

\subsection{Connection to Fluctuation-Dissipation Theorem and Radiation Damping}

The emergence of this $a^{2}\Omega$ term can be anticipated by classical reasoning. The Fluctuation-Dissipation Theorem (FDT), states that for a linear system coupled to a thermal bath, 
  the force $\hat F$ from the bath on the coordinate $\hat x$  has a fluctuation spectrum 
\begin{equation}
\label{FDT}
    S_F  = \mathrm{Im}\left[\chi_x^{-1}\right]\coth\frac{\beta\Omega}{2},
\end{equation}
where $\chi_x^{-1}$ is the inverse of response function of $\hat x$. To be concrete, let us consider an oscillator in an inertial frame with electric charge $q$, which has a Langevin equation of 
\begin{equation}
    m\ddot {\hat x} + m\omega_m^2 {\hat x} + \frac{q^2}{6\pi} \dddot {\hat x} =\hat F  = q\hat E,
\end{equation}
where the triple-derivative term is the Abraham-Lorentz radiation reaction force, and $F$ is the Langevin force from the EM field. Here $\omega_m$ is the mechanical resonant frequency of the oscillator and $m$ is its mass - but regardless of $\omega_m$ and $m$, the EM field's contribution to  $\mathrm{Im}[\chi^{-1}_x]$ is always 
\begin{equation}
    \mathrm{Im}\left[\chi_x^{-1}\right]_{\rm inertial}=\frac{q^2\Omega^3}{6\pi}\,.
\end{equation}
This contribution, together with Eq.~\eqref{FDT}, explains Eq.~\eqref{eq:Wplanck} and fixes its numerical factors. 

If one instead takes the fully relativistic Abraham-Lorentz-Dirac (ALD) force
\begin{equation}
    f^{\mu}=\frac{q^{2}}{6\pi}\left[\frac{d^{3}x^{\mu}}{d\tau^{3}}-\frac{d x^{\mu}}{d\tau}\left(\frac{d^{2}x^{\nu}}{d\tau^{2}}\frac{d^{2}x_{\nu}}{d\tau^{2}}\right)\right]\,.
\end{equation}
and considers a small perturbation of the uniformly accelerated trajectory, $x^{\mu}=\bar{z}^{\mu}+\varepsilon^{\mu}$, with $\ddot{\bar{z}}^{\mu}=a^{2}\bar{z}^{\mu}$ and  $\varepsilon^{\mu}\bar{z}_{\mu}=0$, the ALD formula reduces to
\begin{equation}\label{eq:ALD}
    f^{\mu}=\frac{q^{2}}{6\pi}\left[\frac{d^{3}\varepsilon^{\mu}}{d\tau^{3}}-a^{2}\frac{d \varepsilon^{\mu}}{d\tau}\right]\,.
\end{equation}
This implies enhanced low frequency radiation damping
\begin{equation}
\label{UDWdamping}
    \mathrm{Im}\left[\chi_x^{-1}\right]_{\rm accel.}=\frac{q^2(\Omega^3+a^2\Omega)}{6\pi}\,.
\end{equation}
According to the FDT there is then an enhancement of low frequency thermal fluctuations, Eq.~\eqref{eq:acceleratedpowerspectrum}, and thus the system undergoes the steady decoherence rate Eq.~\eqref{eq:approxdecohrate}.

\begin{figure*}
    \includegraphics[width=0.80\textwidth]{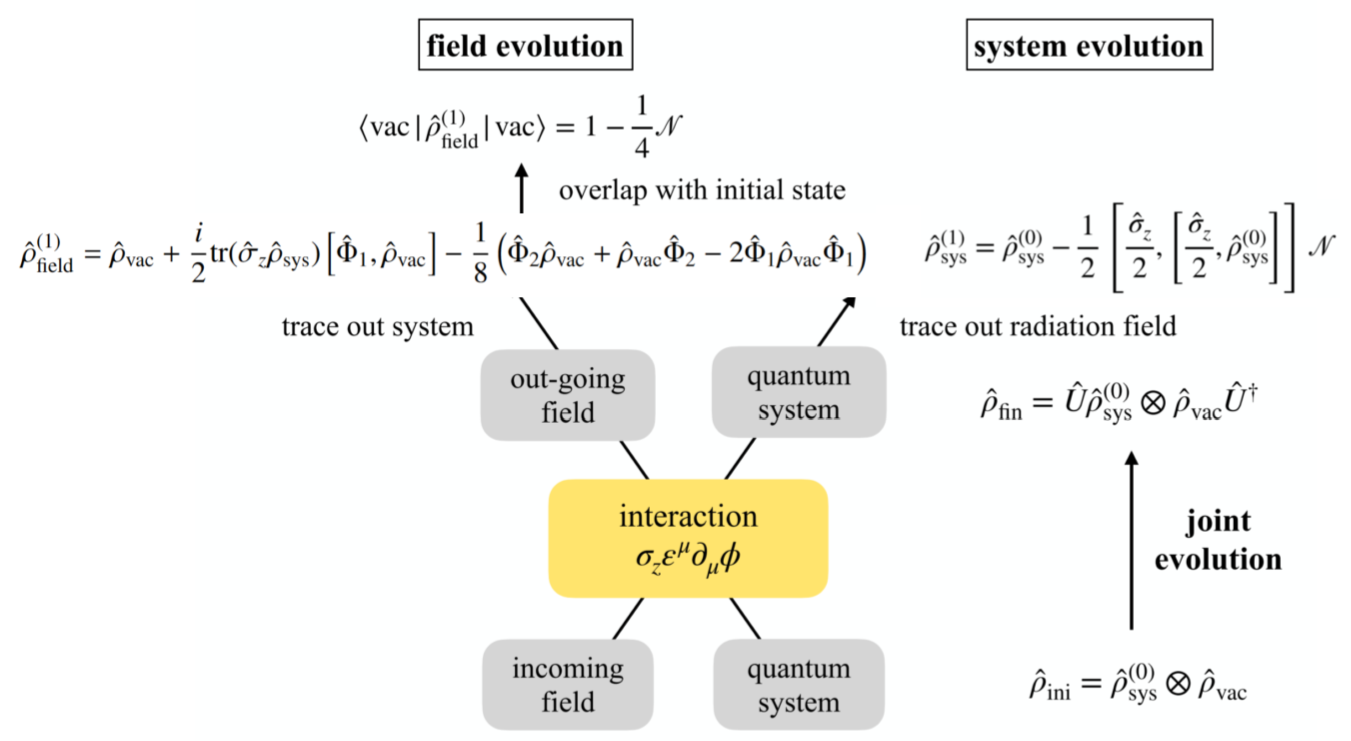}
    \caption{Schematic diagram showing the joint unitary evolution of the field and the system, as well as the evolution of each of their density matrices when the other party is traced out. From this diagram, we can see that the level of decoherence of the quantum system can be extracted consistently by examining the field evolution.  However, in order to obtain all the details of the system's evolution, one must focus on the density-matrix evolution of the system, which means tracing out the outgoing field. Although equations here are written specifically for the dipole calculation in Sec.~\ref{subsec:perturbativedipole}, the structure presented here applies to a broad set of problems.}
    \label{equationdiagram}
\end{figure*}

\subsection{Organization of the paper}

The rest of the paper is organized as follows\footnote{This paper is intended to serve a more pedagogical purpose. To maximize readability across communities we have performed analogous (if not identical) computations in different approaches---separated by section so as to avoid confusion. Additionally, we provide a fair amount of review of known formalism and results.}. In Sec.~\ref{sec:generalformalism} we review general derivations of decoherence, and focus on highlighting a relationship between interpretations of decoherence as arising from radiative losses and from random local forces. As a tool, we map the DSW thought experiment onto a Unruh-DeWitt particle detector model. In Sec.~\ref{sec:thermaldecoherencereview} we perform specific computations of decoherence due to interaction with a thermal bath. We then do detailed computations in Sec.~\ref{sec:emission} of the DSW decoherence, in a scalar field model, by counting the number of photons carrying away ``which-path'' data. In Sec.~\ref{sec:decoherencefunctional} we compute the local force spectrum acting on the particle detector in a number of cases, and show that it matches Gralla and Wei's result for the DSW decoherence rate. In Sec.~\ref{sec:relationtoFDT}  we employ the FDT and discuss the reason why the decoherence rate differs from a standard calculation of thermal decoherence. Finally in Sec.~\ref{sec:conclusions}, we summarize our discussions.

\section{General Formalism of Decoherence}
\label{sec:generalformalism}

\comment{ Let me first start from a simple argument, with a two-oscillator model in the Heisenberg Picture.   One oscillator $(x,p)$ represents the quantum system, while $(Q,P)$ represents the optical field, then the infinitesimal evolution operator is  
\begin{equation}
U = 1 + i \epsilon x P
\end{equation} 
which leads to 
\begin{equation}
\hat x_{\rm out} = \hat x_{\rm in}\,,\quad
\hat p_{\rm out} = \hat p_{\rm in}-\epsilon \hat P_{\rm in}\,,\quad
\end{equation}
and 
\begin{equation}
\hat Q_{\rm out} = \hat Q_{\rm in}\,,\quad
\hat P_{\rm out} = \hat P_{\rm in} + \epsilon \hat x_{\rm in}\,,\quad
\end{equation}
Here ``in'' and ``out'' are used to indicate early- and late-time Heisenberg Operators. The first set of equations represent the oscillator being affected by the incoming field, while the second represents the out-going field being affected by the oscillator.   As we shall see later in this note, the second corresponds to radiation of photons by the oscillator, while the first corresponds to incoming radiation causing decoherence of the oscillator.   Now, the Heisenberg Picture does not offer simple path toward treating a two-level system, therefore we move on to the Schr\"odinger Picture for a dipole approximation, and to the Feynman-Vernon Approach in a general, distributed quantum system. }

In this section, we shall first make a general proof that the decoherence of an Unruh-DeWitt detector can be quantified by the radiation content far from the experiment.  We start with a system coupled to a scalar field, for which we make a dipole approximation, and relate the decoherence between two paths that were first split and recombined to the number of photons radiated by a dipole that first appears and then disappears.  We then do a more general discussion of decoherence, and in the electromagnetic case we make the connection to Wilson loops. 

\subsection{Dipole Approximation}\label{subsec:perturbativedipole}
Suppose our Hamiltonian of the system contains three parts: $\hat H_{\rm field}$ for the free field, $\hat H_{\rm sys}$ that provides the trajectory of the test particle, and the interaction Hamiltonian $\hat V$.  Suppose $\hat H_{\rm sys}$ already makes sure that the wavefunction of the system is given by
\begin{align}
|\psi^{(0)}(t)\rangle &= \left|z^\mu(\tau(t))+\frac{1}{2}\varepsilon^\mu(\tau(t))\right\rangle
+
\left|z^\mu(\tau(t))-\frac{1}{2}\varepsilon^\mu(\tau(t))\right\rangle \nonumber\\
& = |\psi_+ (t) \rangle +|\psi_- (t) \rangle\,.
\end{align}
Here the $|\psi_{\pm}(t)\rangle$ are the two branches, each a very narrow wavepacket; they are mutually separated by $\varepsilon^\mu(\tau)$. 
Note that  along the trajectory, 
\begin{equation}
\dot z^\mu \varepsilon_\mu =0.
\end{equation}
Suppose also that the mass is infinitely heavy, and that the field couples to the position of the mass.  In this situation, the shapes of the packets will not be modified, yet there can be phases added to the two components.  At time $t$, the system's part of the Hilbert space still only involves  $|\psi_{\pm}(t)\rangle$, and we consider the product space between this two-dimensional Hilbert space and that of the field. We essentially have a two-level system - an Unruh-Dewitt detector.

Up to first order, $\hat V$ is given by 
\begin{equation}
\left(
\begin{array}{cc}
\hat\phi(z^\mu+\epsilon^\mu/2) & 0 \\
0 & \hat\phi(z^\mu-\epsilon^\mu/2)
\end{array}\right)
\approx
\hat\phi(z^\mu(\tau))I +\frac{\epsilon^\nu(\tau)}{2}\partial_\nu \hat\phi(z^\mu(\tau)) \hat \sigma_z.
\end{equation}
We throw away the $\hat \phi$ term because it is common to both paths, and gives an overall phaseshift without introducing decoherence.  Therefore, we keep only the second piece, and the evolution operator is given by 
\begin{align}
\hat U&  =T\left\{\exp\left[ i \int d\tau  \frac{\hat\sigma_z}{2} [\varepsilon^\mu  \partial_\mu  \hat\phi]_{\tau}\right]\right\}
\nonumber\\
&= I+  \frac{i\hat\sigma_z}{2} 
\underbrace{\int  d\tau[\varepsilon^\mu  \partial_\mu  \hat\phi]_{\tau}}_{\hat \Phi_1}\nonumber\\
&\quad\;\;\; -  \frac{I}{8} \underbrace{\int_{\tau_2 >\tau_1}  d\tau_2 d\tau_1[\varepsilon^\mu  \partial_\mu  \hat\phi]_{\tau_2}
[\varepsilon^\mu  \partial_\mu  \hat\phi]_{\tau_1}}_{\hat\Phi_2}.
\end{align}
Suppose the system has an initial state described by $\hat \rho_{\rm sys}^{(0)}$, and it forms a product state with the field, $\hat\rho_{\rm ini} = \hat \rho_{\rm sys}^{(0)}\otimes\hat\rho_{\rm vac}$, then the final state is given by 
\begin{align}
\hat\rho_{\rm fin} = \hat \rho_{\rm ini}  & +\frac{i}{2} \left[\hat\sigma_z \hat\Phi_1,\hat\rho_{\rm ini}\right]
\nonumber \\
&-
\frac{1}{8}\left(\hat\Phi_2\hat\rho_{\rm ini}  
 +
\hat\rho_{\rm ini} \hat\Phi_2 - 2\hat\Phi_1 \hat\sigma_z \hat\rho_{\rm ini}\hat\sigma_z \hat\Phi_1
\right).
\end{align} 
Taking the partial trace over the field, we obtain
\begin{equation}
\hat\rho_{\rm sys}^{(1)} = \mathrm{tr}_{\rm field}\hat\rho_{\rm fin} 
=\hat\rho_{\rm sys}^{(0)} -\frac{1}{2}\left[
\frac{\hat\sigma_z}{2} ,\left[\frac{\hat\sigma_z}{2} ,\hat\rho_{\rm sys}^{(0)}\right]\right] \mathcal{N},
\end{equation}
where
\begin{equation}
\mathcal{N} = 
\int d\tau_1 d\tau_2 \varepsilon^\mu (\tau_1)
 \varepsilon^\nu (\tau_2)
\langle 
[ \partial_\mu \hat\phi]_{\tau_1}
[ \partial_\nu \hat \phi]_{\tau_2}
\rangle_{\rm vac}.
\end{equation}
For an initial system state of 
\begin{equation}
\hat\rho_{\rm sys}^{(0)} =\left(
\begin{array}{cc}
\rho_{++} & \rho_{+-}\\
\rho_{-+} & \rho_{--}
\end{array}\right)\,,
\end{equation} 
the final state is given by
\begin{equation}
\label{decohlaw}
\hat\rho_{\rm sys}^{(1)} =\left(
\begin{array}{cc}
\rho_{++} & (1-\tfrac{1}{2}\mathcal{N})\rho_{+-}\\
(1-\tfrac{1}{2}\mathcal{N})\rho_{-+} & \rho_{--}
\end{array}\right).
\end{equation} 
In this way, $\mathcal{N}$ determines the loss in interference contrast and directly measures decoherence.  The formulation here does not depend on the initial system state - we can have any pure or mixed state, as specified by $\rho_{\{++,+-,--\}}$,  and the same law of decoherence \eqref{decohlaw} can be derived (see \cite{Kok:2003udw, Nesterov:2020exl, Xu:2023tdt} for related work).  Note that this law of decoherence is directly caused by incoming field fluctuations.  Depending on the temperature of the surrounding environment, we can have a different decoherence rate. 

However, because of action and reaction, the outgoing radiation  carries information about the interaction between the field and the system.  By examining the outgoing field, one can also determine the decoherence of the system: information lost from the system can be recovered from the outgoing field. By tracing out the system, we have
 \begin{align}
\hat \rho_{\rm field}^{(1)} = &\,\hat\rho_{\rm vac} +\frac{i}{2} \mathrm{tr} (\hat\sigma_z \hat\rho_{\rm sys}) \left[ \hat\Phi_1,\hat\rho_{\rm vac}\right]\nonumber \\
&-
\frac{1}{8}\left( \hat \Phi_2 \hat\rho_{\rm vac} + \hat\rho_{\rm vac}\hat \Phi_2 -2\hat\Phi_1 \hat\rho_{\rm vac}\hat \Phi_1\right)
\end{align} 

which leads to
\begin{equation}
\label{radiated}
\langle \mathrm{vac}| \hat\rho_{\rm field}^{(1)} |\mathrm{vac}\rangle  = 1 -\frac{1}{4}\mathcal{N}.
\end{equation} 
Perturbatively, the final state of the field is in a superposition of zero and one photon states, and Eq.~\eqref{radiated} indicates that the mean number of photons emitted is always $\mathcal{N}/4$.    It is a little curious here that this $\mathcal{N}/4$ does not depend on the quantum state of the system.  A careful examination reveals that this is a special case, because the observable $\sigma_z$ that couples to the radiation field has $\sigma_z^2 =1 $.  The physical interpretation is that whatever the state of the mass, there is always a dipole moment $\hat\sigma_z$, and field amplitude will be $\hat \phi \sim \hat\sigma_z$, with $\hat\phi^2 \sim \sigma_z^2$ always the same.  Also the factor of $1/4$ arises from the fact that each component of the superposition has magnitude $\varepsilon/2$.   Our calculation in this section is illustrated in Figure.~\ref{equationdiagram}.

\subsection{General Split Path Experiment}

The perturbative result of the previous section actually exponentiates. A more general framework for describing decoherence is given by the \emph{decoherence functional}, which itself is a part of the Feynman-Vernon influence functional~\cite{Breuer:2007juk, Feynman:1963fq}. This object generally lives in the integrand of a central system's path-integral, encapsulating all effects due to interactions with the bath system.  However, in situations where the dynamics of the central system are under control, it can often be sufficient to evaluate the influence functional only along the classical trajectory---for the examples considered in this paper, such an approximation is indeed valid. For the two-path system, the reduced density matrix is given by
\begin{equation}
\label{eq:reducedDMIF}
\hat\rho_{\rm sys} =\left(
\begin{array}{cc}
\rho_{++} & \rho_{+-}\,\mathcal{F}[+,-]\\
\rho_{-+}\,\mathcal{F}[-,+] & \rho_{--}
\end{array}\right)\,,
\end{equation} 
where $\mathcal{F}[+,-]$ is the influence functional. It is a functional of the trajectory of the system along its evolution $\psi_{\pm}(t)$.  Only the magnitude is relevant for quantifying decoherence, so one typically writes $|\mathcal{F}|=\exp(-\frac{1}{2}\mathbb{D})$, where $\mathbb{D}\geq0$ is the decoherence functional.

A general review of the influence functional is given in Appendix \ref{sec:influencefunctional}, here we take just a specific application of the general result: for an interaction described by the action
\begin{equation}
    S_{\rm int}[q,\phi]=\sum_{a}\int d\tau\, Q^{a}(q(\tau))\phi_a(\tau)\,,
\end{equation}
where $a$ is a generic, possibly continuous index, $q$ is a set of system variables,  $Q$ is a function of these system variables, and $\phi_{a}$ is a set of bath variables, the decoherence functional is given by
\begin{align}\label{eq:generaldecoherencefunc}
&\mathbb{D}[Q^{+},Q^{-}]=\frac{1}{2}\int^{t}_{-\infty}d\tau\int^{t}_{-\infty}d\tau'\,\bigg\langle\{\phi_{a}(\tau),\phi_{b}(\tau')\}\bigg\rangle_{\rm conn.} \nonumber \\
&\times\Big(Q^{+\,a}(q(\tau))-Q^{-\,a}(q(\tau))\Big)\Big(Q^{+\,b}(q(\tau'))-Q^{-\,b}(q(\tau'))\Big)+\cdots\,,
\end{align}
where the ellipsis denotes connected $(n>2)$-point functions of $\phi$. The notation $Q^{\pm}$ indicates the value of the function $Q(q)$ along the $(\pm)$ trajectory.  The correlation function is evaluated in the initial state of the bath, where it was assumed to be decoupled from the central system. No assumption was made about the state of the bath, nor its composition (aside from it being described by bosonic variables). However, the above expression is only computationally convenient when higher connected $n$-point functions are suppressed. This is certainly the case for photon and graviton baths at low energies.

One of the wonderful pieces of intuition developed by Feynman and Vernon is that decoherence functionals of the form Eq.~\eqref{eq:generaldecoherencefunc} always have an interpretation as arising from the coupling of $Q^{a}$ to a \emph{classical} stochastic driving force $f_{a}$. Indeed, one can disregard any notion of a dynamical bath and simply consider a classical source of the form
\begin{equation}\label{eq:stochasticdecoh}
    S_{\rm source}=\sum_{a}\int d\tau\, Q^{a}(q(\tau))f_{a}(\tau)\,.
\end{equation}
 If $f_{a}$ is a zero mean Gaussian stochastic process with auto-correlator matching the bath's so-called, ``noise kernel'', $\overline{f_{a}(\tau)f_{b}(\tau')} = \Big\langle\{\phi_{a}(\tau),\phi_{b}(\tau')\}\Big\rangle_{\rm conn.}$, then the decoherence it causes on the central system is described by Eq.~\eqref{eq:generaldecoherencefunc}. Thus, this leading order perturbative decoherence always has an interpretation in terms of random local classical forces.

Despite the interpretation of the decoherence functional as arising from random local forces, consistency with Eq.~\eqref{decohlaw} implies that it agrees with the photon count, $\mathbb{D}=\langle\mathcal{N}\rangle$. We will use this as a cross-check in later sections, computing the decoherence both by counting photon flux and by computing the force spectrum on the quantum system.

\subsubsection{Charged Particle Example}

As a concrete example of a decoherence functional consider a minimally coupled electric charge,
\begin{equation}
    S_{\rm int}=q\int d\tau\,\frac{dz^{\mu}}{d\tau}A_{\mu}(z(\tau))\,.
\end{equation}
The thought experiment we'll consider here will be creation of a two-state spatial superposition of the charged particle, where each branch of the wavefunction follows a path $z_{L}(\tau)$ or $z_{R}(\tau)$. We assume that there is a lab trajectory  $\bar{z}(\tau)$ such that at early and late times the trajectories coincide $z_{L}=z_{R}=\bar{z}$. 

The coherence of the evolution can be probed via an interference experiment. A simple example of this involves attaching a qubit degree of freedom to the particle which itself does not couple directly to anything. We assume that the initial state of the qubit is $|\psi\rangle=\alpha|0\rangle+\beta|1\rangle$, and in a manner resembling Stern-Gerlach, the particle follows trajectory $z_{L}$ if the qubit is in state $|0\rangle$ and $z_{R}$ is the qubit is in state $|1\rangle$. While the particle trajectories coincide at late times, the qubit density matrix takes a form entirely analogous to Eq.~\eqref{eq:reducedDMIF}, with the relevant decoherence functional~\cite{Ford:1997yf}
\begin{equation}
    \mathbb{D}[z_{L},z_{R}]=\frac{q^{2}}{2}\int d^{4}x\int d^{4}y\,J_{q}^{\mu}(x)J_{q}^{\nu}(y)\,\left\langle \{A_{\mu}(x), A_{\nu}(y)\}\right\rangle,
\end{equation}
with effective `quantum' source
\begin{equation}
    J_{q}^{\mu}(x)=\int_{-\infty}^{\infty}d\tau\,\left(\dot{z}_{L}^{\mu}(\tau)\delta(x-z_{L}(\tau))-\dot{z}_{R}^{\mu}(\tau)\delta(x-z_{R}(\tau))\right).
\end{equation}
Note that since the quantum source vanishes at early and late times, a simple integration by parts proves that it is conserved, $\partial_{\mu}J^{\mu}_{q}=0$. Since the source is conserved, the decoherence functional is independent of the gauge choice for $A_{\mu}$. 

This gauge independence can be manifested by writing this integral in the language of differential forms
\begin{equation}\label{eq:decoherencefunctionalloopintegral}
    \mathbb{D}[z_{L},z_{R}]=q^{2}\left\langle\oint_{\mathcal{C}} A\oint_{\mathcal{C}} A\right\rangle=q^{2}\left\langle\int_{\mathcal{S}} F\int_{\mathcal{S}} F\right\rangle\,,
\end{equation}
where the curve $\mathcal{C}=z_L*(-z_R)$ is the closed loop going forward along $z_L$ and then back along $z_R$, and $\mathcal{S}$ is any two-dimensional surface anchored to $\mathcal{C}$. Note that although we've omitted the anti-commutator in the correlation function, the integrand is symmetric, so we will only be picking out the symmetric part of the Wightman function above. This geometric picture goes a step further if we observe that the magnitude of the whole influence functional may then be written as the Wilson loop
\begin{equation}
    |\mathcal{F}[z_{L},z_{R}]|=\left\langle \exp\left(iq\oint_{\mathcal{C}} A\right)\right\rangle\,.
\end{equation}

\subsubsection{Relevant Models}

We are interested in long-time persistent effects, for which the relevant field modes are very long wavelength. Small scale details of experiments conducted in a finite size laboratory are irrelevant to these modes. Their coupling to everything in the lab is best modelled by an effective worldline description where the field couples to localized dynamical multipole-moment operators which characterize the dynamics of the experiment, as in e.g.~\cite{Goldberger:2004jt,Goldberger:2009qd,Ross:2012fc}.

For present calculations we'll consider three effective models. In each case we will consider a two-state system carried along a lab worldline $\bar{z}^{\mu}(\tau)$ and coupled to a quantum field. We will consider: \textit{a)} the original Unruh-DeWitt model of a monopole-moment coupling to a massless scalar field, \textit{b)} a dipole-moment coupling to a massless scalar field, and \textit{c)} an electric-dipole coupling to the electromagnetic field. 

\paragraph{Scalar monopole:}
We have a Hamiltonian
\begin{equation}
    H_{\rm int}(\tau)=\lambda(\tau) \sigma_{z} \phi(z(\tau))\,,
\end{equation}
where $\lambda$ is a time dependent coupling, $\sigma_{z}=\pm1$ is the detector's monopole moment operator, and we'll assume that the detector has no other dynamics (eg. no bias terms such as $H_{0}=-\vec{B}\cdot\vec{\sigma}$). We will assume that $\lambda$ has compact support, and that observations on the detector are performed after decoupling. The decoherence functional for this particle detector system is then simply
\begin{equation}\label{eq:monopoledecohfunc}
    \mathbb{D}=2\int^{\infty}_{-\infty}d\tau\int^{\infty}_{-\infty}d\tau'\,\lambda(\tau)\lambda(\tau')\Big\langle\{\phi(z(\tau)),\phi(z(\tau'))\}\Big\rangle\,.
\end{equation}

\paragraph{Scalar dipole:}
The action is
\begin{equation}
    S_{\rm int}=\int d\tau\, d^{\mu}(\tau)\partial_{\mu}\phi(\bar{z}(\tau))\,.
\end{equation}
The dipole moment $d^{\mu}$ can be taken without loss of generality as defined in the local orthonormal frame, $d^{\mu}(\tau)=d^{I}(\tau)e_{I}^{\mu}(\tau)$, where $e_{I}^{\mu}(\tau)\dot{\bar{z}}_{\mu}(\tau)=0$. For a superposition in which the dipole moment is either $d^{I}(\tau)=\pm\tfrac{1}{2}q\varepsilon^{I}(\tau)$, we have the decoherence functional
\begin{equation}\label{eq:scalardipoledecohfunc}
    \mathbb{D}=\frac{q^{2}}{2}\int^{\infty}_{-\infty}d\tau d\tau'\,\varepsilon^{\mu}(\tau)\varepsilon^{\nu}(\tau')\Big \langle\{\partial_{\mu}\phi(\bar{z}(\tau)),\partial_{\nu}\phi(\bar{z}(\tau'))\}\Big\rangle\,.
\end{equation}

\paragraph{Electric dipole:}
The action is
\begin{equation}
    S_{\rm int}=q\int d\tau\, \varepsilon^{\mu}(\tau)\dot{\bar{z}}^{\nu}(\tau)F_{\mu\nu}(\bar{z}(\tau))\,,
\end{equation}
for a dipole moment $q\varepsilon^{\mu}$. The dipole moment may again be taken as orthogonal to the four-velocity, and we consider a superposition of the following trajectories, $\varepsilon^{\mu}(\tau)=\pm\tfrac{1}{2}\varepsilon^{I}(\tau)e_{I}^{\mu}(\tau)$. The decoherence functional is then
\begin{equation}\label{eq:elecdipoleDF}
    \mathbb{D}=\frac{q^{2}}{2}\int_{-\infty}^{\infty} d\tau d\tau'\,\varepsilon^{I}(\tau)\varepsilon^{J}(\tau)\, \left\langle \{E_{I}(\tau),E_{J}(\tau')\}\right\rangle\,,
\end{equation}
where the local electric field is defined as $E_{I}(\tau)=e_{I}^{\mu}(\tau)\dot{\bar{z}}^{\mu}(\tau)F_{\nu\mu}(\bar{z}(\tau))$. The open quantum dynamics of radiation coupled dipoles has been studied extensively in the non-relativistic limit, see e.g.~\cite{Barone:1991zz,Breuer:2007juk} and refs. therein, however Eq.~\eqref{eq:elecdipoleDF} as presented here is valid for general trajectories and is not limited to Minkowski spacetime.

In Sec.~\ref{sec:emission}, we will focus on the scalar dipole in a uniformly accelerated lab, as the simplest model which captures the physics observed by DSW. We will compute $\mathcal{N}$ in two ways, one using the characteristic approach taken by DSW, and another using the standard Cauchy approach. In Sec.~\ref{sec:decoherencefunctional} we will compute $\mathbb{D}$ for the scalar and electric dipoles, finding exact agreement with the previous computations.

\begin{figure}
    \includegraphics[width=0.45\textwidth]{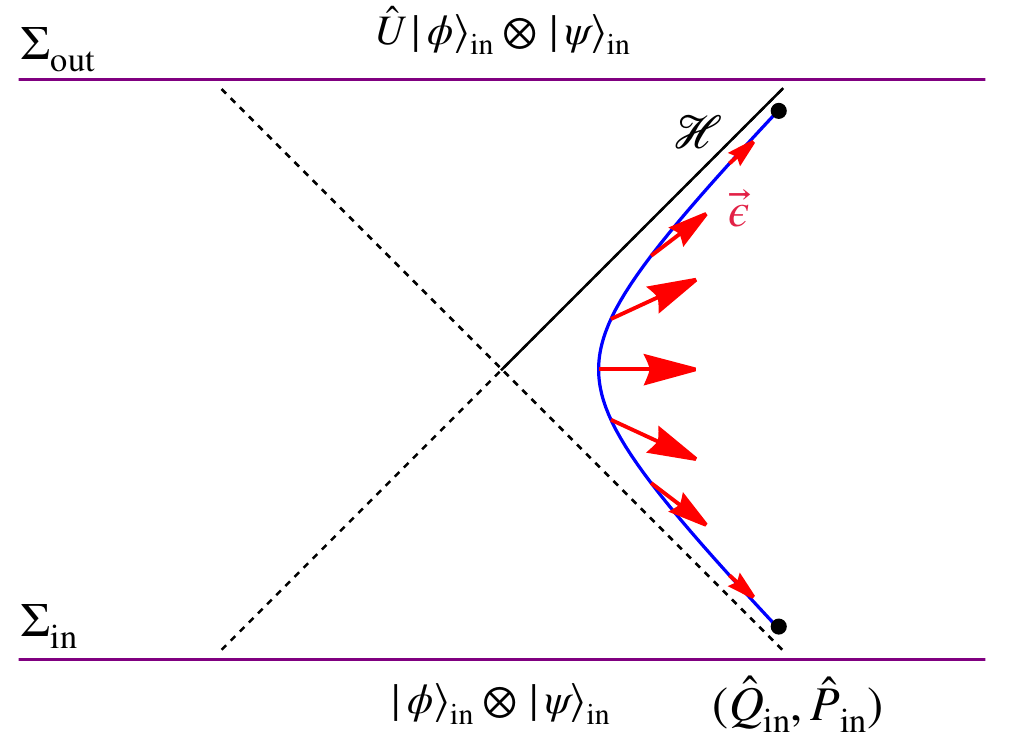}
    \caption{Space-time diagram illustrating the trajectory (blue) of a constantly accelerating quantum system and the deviation vector $\vec \varepsilon$ (red arrows) along which a superposition state is formed.  We have shown an initial Cauchy surface $\Sigma_{\rm in}$, from which incoming radiation originates, and a final Cauchy surface $\Sigma_{\rm out}$, from which outgoing radiation exits. We have also shown on the figure a Killing Horizon $\mathcal{H}$, on which outgoing radiation can be studied.}
\end{figure}

\section{Standard Thermal Decoherence}\label{sec:thermaldecoherencereview}

Decoherence of a system due to interactions with a thermal bath is a well studied phenomenon~\cite{Breuer:2007juk}. In this section we will focus on the case of simple Unruh-DeWitt particle detectors at rest, coupled to a scalar field at finite temperature. For the simplest case, a qubit with monopole moment coupling, thermal decoherence proceeds with a steady rate. The analogous computation in an accelerated frame gives exactly the same result, provided that $a=2\pi\beta^{-1}$. This is a classic example~\cite{Unruh:1976db,Birrell:1982ix}, but is useful to review before we proceed.  Interestingly though, when we extend to the case of a dipolar coupling we'll find that there is zero thermal decoherence\footnote{Here, and throughout, we are referring only to the linear dependence of $\mathbb{D}$ on the total proper time $T$. There is generically also $\log(T)$ contributions to the decoherence functional coming from radiation which reaches null infinity~\cite{Wilson-Gerow:2018egh,DeLisle:2022pjo, Neuenfeld:2018fdw,Gralla:2023oya}, which we are not investigating in this work.}. This is consistent with expectations coming from the vanishing Thompson cross-section in the limit of no recoil, and bolsters the confusion around the local explanation for the DSW soft photon decoherence. 

We'll start with Eq.~\eqref{eq:monopoledecohfunc} for the decoherence of a monopole-moment coupled detector. In the vacuum state the symmetric correlation function is given by
\begin{equation}\label{eq:freescalarG1}
    G^{(1)}(x,y)=\langle 0|\{\phi(x),\phi(y)\}|0\rangle=\frac{1}{2\pi^{2}}\frac{1}{(x-y)^{2}},
\end{equation}
where we are using mostly-plus Lorentzian signature, and the principal value is implied.

Consider first an inertial lab trajectory, $z^{\mu}(\tau)=u^{\mu}\tau$, and states of the bath such that this correlation function is stationary, i.e. a function only of the combination $\tau-\tau'$. Since $\lambda$ is equal to a constant value $\lambda_{0}$ for a long time, we effectively have a constant decoherence rate, $\frac{1}{2}\mathbb{D}\approx\Gamma T$, with
\begin{equation}
    \Gamma=\lambda_{0}^{2}\int_{-\infty}^{\infty}d\tau \,\Big\langle\{\phi(u\tau),\phi(0)\}\Big\rangle\,. 
\end{equation}
 Concretely we may then write the vacuum decoherence rate as
\begin{align}
     \Gamma_{\rm inertial}=\frac{-\lambda_{0}^{2}}{2\pi^{2}}\int_{-\infty}^{\infty}d\tau \,\frac{1}{(\tau-i\epsilon)^{2}}=0\,.
\end{align}
This vanishing can be seen simply by closing the contour in the lower half plane. 

Consider now the inertial observer at rest relative to a thermal bath with inverse temperature $\beta$. For systems in thermal equilibrium we have the KMS relation
\begin{equation}
    \textrm{Tr}\left(e^{-\beta H}\,\phi(x^{0},\vec{x})\phi(0,0)\right)=\textrm{Tr}\left(e^{-\beta H}\,\phi(0,0)\phi(x^{0}+i\beta,\vec{x})\right)\,,
\end{equation}
which implies the periodicity of the symmetric correlation function in imaginary time,
\begin{equation}\label{eq:imagesum}
    G^{(1)}_{\beta}(x^{0},\vec{x})= G^{(1)}_{\beta}(x^{0}+i\beta,\vec{x})\,.
\end{equation}
Together with the equation of motion, this periodicity condition implies that the free field thermal correlation function is given by a simple image-sum over the vacuum correlation function
\begin{align}
    G^{(1)}_{\beta}(x^{0},\vec{x})&= \sum_{l=-\infty}^{\infty}\,G^{(1)}(x^{0}+il\beta,\vec{x})\,.
    \end{align}
For the free scalar field this is
\begin{equation}
    \Big\langle\{\phi(u\tau),\phi(0)\}\Big\rangle_{\beta}=\sum_{l=-\infty}^{\infty}\frac{-1}{2\pi^{2}}\frac{1}{(\tau+il\beta)^{2}}\,.
\end{equation}
    
We can then use the following identity 
\begin{equation}
    \frac{1}{\sin^{2}(\pi x)}=\frac{1}{\pi^{2}}\sum_{l=-\infty}^{\infty}\frac{1}{(x-l)^{2}}\,,
\end{equation}
to obtain the thermal correlation function
\begin{equation}\label{eq:thermalG1}
    G^{(1)}_{\beta}(\tau,0)=\frac{-1}{2\beta^{2}}\frac{1}{\sinh^{2}\left(\frac{\pi\tau}{\beta}\right)}\,.
\end{equation}
Integrating over $\tau$, the corresponding decoherence rate is
\begin{equation}
    \Gamma_{\beta}=\frac{\lambda_{0}^{2}}{\pi\beta}\,.
\end{equation}

Compare this with the experience of a uniformly accelerated detector moving through a field in the Minkowski vacuum state. For the trajectory
\begin{equation}
\bar{z}^\mu (\tau) = (t,x,y,z) =  \left[\frac{1}{a}\sinh (a\tau),0,0,\frac{1}{a}\cosh (a\tau)\right]\,,
\end{equation} 
the symmetric correlator Eq.~\eqref{eq:freescalarG1} evaluated along the worldline is
\begin{equation}\label{eq:acceleratedG1}
    G^{(1)}(z(\tau),z(0))=\frac{-a^{2}}{8\pi^{2}}\frac{1}{\sinh^{2}\left(\frac{a\tau}{2}\right)}\,.
\end{equation}

The fact that the correlation functions Eq.~\eqref{eq:acceleratedG1} and Eq.~\eqref{eq:thermalG1} are identical under the identification of $\beta=2\pi/a$, is a statement of the Unruh effect. We then see that for a uniformly accelerated detector with monopolar coupling to the scalar field the decoherence can be completely understood \emph{in the local frame} as coming from interaction with a thermal bath. 

Now consider the decoherence functional for a dipole coupled to a scalar field, Eq.~\eqref{eq:scalardipoledecohfunc}. For a lab at rest relative to a thermal bath, the correlation function is stationary and we have the decoherence rate
\begin{equation}
    \Gamma=\frac{q^{2}\varepsilon^{i}\varepsilon^{j}}{4}\int^{\infty}_{-\infty}d\tau \,\Big \langle\{\partial_{i}\phi(u\tau),\partial_{j}\phi(0)\}\Big\rangle_{\beta}\,.
\end{equation}
Taking the appropriate derivatives of Eq.~\eqref{eq:freescalarG1}, this correlation function is then written as the thermal image sum
\begin{equation}
    \Big \langle\{\partial_{i}\phi(u\tau),\partial_{j}\phi(0)\}\Big\rangle_{\beta}=\frac{\delta_{ij}}{\pi^{2}}\sum_{l=-\infty}^{\infty}\frac{1}{(\tau+il\beta)^4}\,.
\end{equation}
Carefully evaluating the integral over $\tau$, one finds for $n\geq1$ 
\begin{equation}\label{eq:tauaverage}
   \int d\tau \sum_{l=-\infty}^{\infty}\,\frac{1}{(\tau+il\beta)^{2n}}=\begin{cases}
   \frac{-2\pi}{\beta },& \text{if } n=1\\
    0,              & \text{otherwise}
\end{cases}\,.
\end{equation}
Evidently, while the monopole-moment coupled detector $(n=1)$ demonstrated thermal decoherence, the dipole and higher derivative couplings have vanishing thermal decoherence rates. 

The DSW thought experiment can be mapped onto an Unruh-DeWitt detector computation, albeit one with dipolar coupling. Given that the monopole-coupled Unruh-DeWitt decoherence was entirely explained by thermal fluctuations, one would naturally expect that the DSW decoherence would also be explained by thermal fluctuations. This expectation is challenged, however, by the above explicit computation.

Next, in Sec.~\ref{sec:emission}, we will compute decoherence via photon emission, while in Sec.~\ref{sec:decoherencefunctional}, we will perform an alternative computation where we carefully track the effect of the field's vacuum fluctuations on the detector, achieving equivalent results and addressing the apparent inconsistency between those results and the ones in this section.

\section{Photon Emission Rate of Accelerated Quantum Systems (DSW)}
\label{sec:emission}
In this section, we shall compute the photon emission rate $\mathcal{N}$ of a uniformly accelerated dipole using two different approaches: the characteristic approach of DSW and a Cauchy approach. We have generalized the DSW treatment by considering a dipole along a generic direction. The results of the two approaches will agree with each other.

\subsection{Characteristic Calculation}
Suppose we have a trajectory 
\begin{equation}\label{eq:acceleratedtrajectory}
z^\mu (\tau) = (t,x,y,z) =  \left[\frac{1}{a}\sinh (a\tau),0,0,\frac{1}{a}\cosh (a\tau)\right],
\end{equation} 
where $\tau$ is the proper time and $a$ the proper acceleration.
On this trajectory, we have a particle with dipole moment
\begin{equation}
q\varepsilon^\mu(\tau) = \frac{q}{a}   \left[ \varepsilon_Z(\tau) \sinh (a\tau),\varepsilon_X(\tau),\varepsilon_Y(\tau),\varepsilon_Z(\tau)\cosh (a\tau) \right].
\end{equation}

Here $\varepsilon_{X,Y,Z}$ are components of the dipole moment along the $x$, $y$ and $z$ directions. Note that for $\varepsilon_Z$, the direction of $\varepsilon^\mu$ contains both $t$ and $z$ components, so $\varepsilon^\mu$ can maintain orthogonality with $\dot z^\mu(\tau)$.

It is understandable that $\varepsilon$ will become very close to zero within a finite duration.  In this way, the dipole source does not extend to infinity.  Let us consider the scalar field sourced by this dipole.  We consider a new trajectory, with
\begin{equation}
z^\mu (\tau,\lambda ) = z^\mu(\tau) + \lambda \varepsilon^\mu(\tau),
\end{equation}
and four-velocity vector 
\begin{equation}
u^\mu (\tau,\lambda ) = \dot z^\mu(\tau) + \lambda \dot \varepsilon^\mu(\tau), 
\end{equation}
where $\lambda = \pm \frac{1}{2}$.
For any field point, the Lienard-Wiechert potential is given by
\begin{equation}
\phi (\vec x)= \frac{\delta}{\delta\lambda}  \left[- \frac{q}{4\pi} \frac{1}{[x^\mu - z^\mu (\tau_{\rm ret}(\vec x,\lambda),\lambda)] u_\mu (\tau_{\rm ret}(\vec x,\lambda))}\right],
\end{equation} 
where $\tau_{\rm ret}(\vec x,\lambda)$ satisfies
\begin{equation}
\left[x^\mu - z^\mu (\tau_{\rm ret}(\vec x,\lambda),\lambda)\right] \left[ x_\mu - z_\mu (\tau_{\rm ret}(\vec x,\lambda),\lambda)\right] =0,
\end{equation}
and $\vec u (\tau)$ is the 4-velocity of the particle.  More specifically, consider a point 
\begin{equation}
P = (\xi,x,y,\xi)
\end{equation}
on the Rindler horizon $\mathcal{H}$ of the particle. Here $\xi$ is the retarded time on the horizon.  Relating to $\xi$,  the retarded proper time on the trajectory  is given  by 
\begin{align}
&\tau_{\rm ret}(\xi,x,y) \nonumber\\
=&\frac{1}{a}\log\frac{2a\xi}{1+a^2\rho^2}  \nonumber\\
+&\frac{\lambda }{a}\left[\frac{2a x \varepsilon_X(\tau) +2a y\varepsilon_Y(\tau) +[-1+a^2\rho^2]\varepsilon_Z(\tau)}{1+a^2\rho^2}\right]_{\tau = \tau_{\rm ret}^{(0)}(\xi,x,y)},
\end{align}
where we have defined
\begin{equation}
\rho = x^2+y^2\,,\quad \tau_{\rm ret}^{(0)}(\xi,x,y) = \frac{1}{a}\log\frac{2a\xi}{1+a^2\rho^2}. 
\end{equation} 
This allows us to compute the dipole potential for $\xi>0$, which is given by 
\begin{widetext}
\begin{align}
\phi(P) =\left\{
-\frac{qa}{2\pi(1+a^2\rho^2)^2}
 \left [
[ 2a \varepsilon_Z +(1- a^2\rho^2)\varepsilon_Z'] -
2a x (a \varepsilon_X +  \varepsilon_X ') -
2a  y (a \varepsilon_Y +  \varepsilon_Y ') \right]
 \right\}_{\tau = \tau_{\rm ret}^{(0)}(\xi,x,y)}.
\end{align} 
From here, we will see that a constant $\phi$ exists for the duration in $\xi$ which tracks back to a retarded time $\tau_{\rm ret}$ that had a non-zero dipole.  It was from here that Danielson, Satishchandran and Wald argued for the decoherence proportional to proper time.  We can go on and compute the radiative flux through the Rindler horizon.

Note that this is only non-zero for $\xi>0$. 
We can then write
\begin{align}
\label{phixi}
&\phi(\omega,x,y) \nonumber\\
\equiv &\int_0^{+\infty} d\xi \,e^{i\omega\xi} \phi(\xi,x,y,\xi)  
\nonumber\\
=&  -\frac{qa^2}{2\pi }\int_{-\infty}^{+\infty} d\tau 
\, 
\frac{[\varepsilon_Z(\tau)+\frac{1-a^2\rho^2}{2a}\varepsilon'_Z(\tau)]  -  x[a \varepsilon_X(\tau)+\varepsilon_X'(\tau)]- y[a \varepsilon_Y(\tau)+\varepsilon_Y'(\tau)]}{(1+a^2\rho^2)}
\exp\left[a\tau+i\omega\frac{1+a^2\rho^2}{2a}e^{a\tau}\right]. 
\end{align} 
\end{widetext}
The photon number is then given by (see also Appendix~\ref{sec:jacobian})
\begin{align}
\label{Neq}
\mathcal{N} &=  \iint_{-\infty}^{+\infty} dxdy \int_0^{+\infty}\frac{d\omega}{2\pi}2\omega  |\phi(\omega,\rho)|^2. 
\end{align} 
To compute the photon number radiated, we first perform an integral by parts in Eq.~\eqref{phixi}, obtaining an integral only over $\varepsilon_{x,y,z}(\tau)$ and not its derivatives, before inserting into Eq.~\eqref{Neq}.  This leads to 
\begin{equation}
    \mathcal{N} = \iint d\tau_1 d\tau_2 W_{\rm DSW}^{IJ} (\tau_1,\tau_2)\varepsilon_I^*(\tau_1) \varepsilon_J(\tau_2).
\end{equation}
Here we have allowed $\varepsilon$ to be complex, for purposes that will become clear later.  We find
\begin{align}\label{eq:Nscalardipole}
    W_{\rm DSW}^{xx,yy}(\tau_1,\tau_2)&=\frac{q^2a^4}{32\pi^2}\left(\frac{1}{-i\epsilon + \sinh\frac{a(\tau_1-\tau_2)}{2}}\right)^4, \nonumber\\
        W_{\rm DSW}^{zz} &=
\frac{q^2a^4}{32\pi^2}\left(\frac{1}{-i\epsilon + \sinh\frac{a(\tau_1-\tau_2)}{2}}\right)^4\nonumber\\
&-
\frac{q^2a^4}{16\pi^2}\left(\frac{1}{-i\epsilon + \sinh\frac{a(\tau_1-\tau_2)}{2}}\right)^2.
\end{align}
Using 
\begin{equation}
    \varepsilon_J(\tau) = \int \frac{d\Omega}{2\pi} \tilde \varepsilon_J(\Omega)e^{-i\Omega \tau},
\end{equation}
we can write 
\begin{equation}
\mathcal{N} = \int_{-\infty}^{+\infty}W^{IJ}_{\rm DSW}(\Omega) \tilde\varepsilon^*_I(\Omega)
\tilde\varepsilon_J(\Omega) \frac{d\Omega}{2\pi}
\label{eqNdsw}
\end{equation} 
and 
\begin{align}\label{eq:DSWwightman}
    W_{\rm DSW}^{xx}(\Omega )& =
    \frac{q^2a^2\Omega}{12\pi}\left[1+\left(\frac{\Omega}{a}\right)^2\right]
    \left(\coth\frac{\pi\Omega}{a}+1\right) \nonumber\\  
    W_{\rm DSW}^{zz}(\Omega )& =
W_{\rm DSW}^{xx}(\Omega )+\frac{q^2a^2\Omega}{4\pi} \left(\coth\frac{\pi\Omega}{a}+1\right). 
\end{align} 
To perform the Fourier transform here, one way is to make a substitution $ z = e^{a(\tau_1-\tau_2)}$  and $\tau_1 -\tau_2 = a^{-1}\log z$ and perform a contour integral that goes below the positive real axis, turns above the real axis, and then around a large circle.  Here the branch-cut for $\log$ should start from the origin and run along the positive real axis.

We notice that $W$ is complex in the time domain, therefore in the frequency domain, we have $W(\Omega) \neq W^*(-\Omega)$. 
However, $W$ is still a Hermitian operator on $\varepsilon$, with
\begin{equation}
    W(\tau_1,\tau_2)=    W^*(\tau_2,\tau_1)\,,
\end{equation}
which guarantees that $\mathcal{N}$ is real-valued.

To compute the number of photons emitted by the dipole, we will restrict to real-valued $\varepsilon$ in the time domain.  Correspondingly, we need to define a symmetrized force spectrum 
\begin{equation}
    S_F(\Omega) =\frac{W(\Omega)+W(-\Omega)}{2},  
\end{equation}
which in our case is 
\begin{align}
    S_F^{xx,yy}(\Omega) &= \frac{q^2a^2\Omega}{12\pi}\left[1+\left(\frac{\Omega}{a}\right)^2\right]
    \coth\frac{\pi\Omega}{a}, \\
    S_F^{zz}(\Omega)& =   \frac{q^2a^2\Omega}{12\pi}\left[4+\left(\frac{\Omega}{a}\right)^2\right]
    \coth\frac{\pi\Omega}{a}.
\end{align}
This will indeed relate to the force on the dipole as we explicitly compute its decoherence in Sec.~\ref{sec:decoherencefunctional}.   The fact that $S_F(\Omega) \sim a^3$ as $\Omega\rightarrow 0$ leads to the scaling of 
\begin{equation}
\mathcal{N} \sim q^2 a^3  |\varepsilon|^2 T.
\end{equation} 
More specifically, suppose $\varepsilon(\tau)$ is a constant value $\varepsilon$ for a long (proper time) duration $T$ and smoothly transitions to zero at both ends.  If $T \gg 1/a$ and $\varepsilon$ only has low-frequency components at $\Omega a \ll 1$, we can write
\begin{align}\label{eq:charradiatedphotonnum}
\mathcal{N}=&\int_{-\infty}^{+\infty}  
  S_F^{IJ}(\Omega)
  \tilde\varepsilon_I^*(\Omega)\tilde\varepsilon_J(\Omega) 
  \frac{d\Omega}{2\pi}\nonumber\\ 
\approx & S_F^{IJ}(0) \int_{-\infty}^{+\infty}  
  \tilde\varepsilon_I^*(\Omega)\tilde\varepsilon_J(\Omega) 
  \frac{d\Omega}{2\pi}\nonumber\\=&
 S_F^{IJ}(0)  \int_{-\infty}^{+\infty} \varepsilon_I(\tau) \varepsilon_J(\tau)  d\tau = \varepsilon_I \varepsilon_J  T S_F^{IJ}(0) \nonumber \\
 =&\frac{Tq^{2}a^{3}}{12\pi^{2}}\left(|\varepsilon|^{2}+3\varepsilon_{Z}^{2}\right)\,.
\end{align}
In the second line, the narrow bandwidth of $\tilde\varepsilon$ limits the integration bound to the region where $S_F(\Omega)$ is well approximated by $S_F(0)$, while in the third line we have applied the Parseval theorem.

\subsection{Cauchy Calculation}

It is more conventional to compute the photon number radiated by using the Fourier approach. For a particle trajectory $z^\mu(\tau)$ and dipole trajectory $\varepsilon^\mu(\tau)$, we can use a classical scalar field equation 
\begin{equation}
\square\phi = \int d\tau \varepsilon^\nu(\tau) \partial_\nu \delta^{(4)}(x^\mu - z^\mu(\tau)).
\end{equation}
This can be solved using a Fourier transform, with 
\begin{equation}
\label{intphi}
\phi (t,\mathbf{x}) =\int \frac{d\omega}{2\pi }e^{-i\omega t } \int \frac{d^3\mathbf{k}}{(2\pi)^3} e^{-i \mathbf{k}\cdot\mathbf{x}} \frac{S(k^\mu) }{(\omega+i\epsilon)^2 - \mathbf{k}^2 }.
\end{equation} 
Where the $i\epsilon$ prescription above selects the retarded boundary conditions. The source term $S$ is given as an integral along the path of the system,  
\begin{equation}
S(k^\mu)=\int_{-\infty}^{+\infty} d\tau  \, ik_\mu\varepsilon^\mu(\tau)  e^{-i k_\mu z^\mu(\tau)}.
\end{equation}
We note that $k^\mu =(\omega,\mathbf{k})$ does not yet have to be on-shell in the above integrals, which take place in the entire 4-dimensional $k^\mu$ space. 

Since the source term is practically limited within a finite duration in time, the integral over $\omega$ in Eq.~\eqref{intphi} can be done by extracting the residue pole at $\omega =|\mathbf{ k}|$ , leading to 
\begin{equation}
\left[\phi(t,\mathbf{x})\right]_{\rm late} =q\int \frac{d^3 \mathbf{k}}{(2\pi)^3} e^{-i\omega_{\mathbf{k}} t}
\frac{1}{2\omega_{\mathbf{k}}} S(\mathbf{k})  e^{i\mathbf{k}\cdot\mathbf{x}}.
\end{equation} 
Evaluating the integral at late times allow us to use $S(\mathbf{k})$, which is the on-shell value of $S(k^\mu)$. In this way, the radiated number of photons will be given by 
\begin{equation}
\mathcal{N}=q^2\int \frac{d^3 \mathbf{k}}{(2\pi)^3} \frac{1}{2\omega_{\mathbf{k}}} |S^2(\mathbf{k})|.
\end{equation}
Specifically, in our case, we can define
 \begin{equation}
\label{defS}
S(\mathbf{k}) = \int_{-\infty}^{+\infty} d\tau s(\mathbf{k},\tau)
\end{equation}
and 
 \begin{align}
 \label{defss}
  s(\mathbf{k},\tau) 
 =&\varepsilon^\mu(\tau)k_\mu e^{i k_\mu z^\mu(\tau)}\nonumber\\
=&
 \Big\{  \varepsilon_Z(\tau)[-\omega_{\mathbf{k}}  \sinh ( a\tau) +k_z \cosh (a\tau)]\nonumber\\
&  k_ x \varepsilon_X(\tau) +k_y \varepsilon_Y(\tau)
\Big\}e^{-i{\frac{\omega_{\mathbf{k}}}{a}[ \sinh (a\tau)  - \cos\theta \cosh(a\tau)]} }\,.
\end{align} 
This leads to exactly the same decoherence rate as in the DSW case. It is interesting to note that evaluating the radiation flux on the just the future horizon, without consideration future null infinity, is sufficient to reproduce the effect.  We can see this from the Appendix.

One tip for obtaining the time-domain Wightman function is to add a regularization term $e^{\epsilon\omega}$, $\epsilon\rightarrow 0+$ in the $\mathbf{k}$ integral, and only compute the case of $\tau_1+\tau_2=0$ --- because $W(\tau_1,\tau_2)$ only depends on $\tau_1-\tau_2$.

\section{Decoherence from vacuum fluctuations (Unruh-DeWitt):  Scalar and Electric Dipoles}
\label{sec:decoherencefunctional}

In this section, we shall employ the Unruh-DeWitt approach to compute decoherence rate due to incoming vacuum fluctuations.

\subsection{Scalar-dipole Superposition Decoherence}

Let us directly compute the decoherence rate for a constantly accelerating system via the noise term in the influence functional. That is, let us evaluate
\begin{equation}\label{decohint}
    \mathbb{D}=\frac{q^{2}}{2}\int^{\infty}_{-\infty}d\tau d\tau'\,\varepsilon^{\mu}(\tau)\varepsilon^{\nu}(\tau')\Big \langle\{\partial_{\mu}\phi(\bar{z}(\tau)),\partial_{\nu}\phi(\bar{z}(\tau'))\}\Big\rangle\,.
\end{equation}
with the field in the Minkowski vacuum state. The correlation function above is the symmetric part of the Wightman function
\begin{equation}
    G^{(+)}(x,y)=\langle 0|\phi(x)\phi(y)|0 \rangle=\frac{-1}{4\pi^{2}}\frac{1}{(x^{0}-y^{0}-i \epsilon)^{2}-(\vec{x}-\vec{y})^{2}}\,.
\end{equation}

Since the Minkowski vacuum is invariant under Lorentz boosts, and points along the uniformly accelerated worldline are all connected by boosts, the pullback of the scalar Wightman function to the worldline is a function of $\tau-\tau'$ alone,
\begin{equation}
    G^{(+)}(\bar{z}(\tau),\bar{z}(\tau'))=\frac{-a^2}{16\pi^{2}}\left(\frac{1}{\sinh\left(\tfrac{a}{2}(\tau-\tau')\right)-i\epsilon}\right)^{2}\,.
\end{equation}
The partial derivatives spoil the stationarity of the Wightman function, however when they are projected into the local orthonormal frame the stationarity is restored:
\begin{align}
     G^{(+)}_{IJ}(\tau,\tau')&\equiv e_{I}^{\mu}(\tau)e_{J}^{\nu}(\tau')\frac{\partial}{\partial x^{\mu}}\frac{\partial}{\partial y^{\nu}}G^{(+)}(x,y)\bigg|_{x=\bar{z}(\tau),y=\bar{z}(\tau')} \nonumber \\
     &=\frac{a^4}{32\pi^{2}}\delta_{IJ}\left(\frac{1}{\sinh\left(\tfrac{a}{2}(\tau-\tau')\right)-i\epsilon}\right)^{4}\nonumber \\
     &-\frac{a^4}{16\pi^{2}}\delta_{I}^{Z}\delta_{J}^{Z}\left(\frac{1}{\sinh\left(\tfrac{a}{2}(\tau-\tau')\right)-i\epsilon}\right)^{2}\,.
\end{align}
Here the vectors $e^{\mu}_{I}$ for $I=(X,Y,Z)$ form a local orthonormal basis orthogonal to $\dot{\bar{z}}^{\mu}$. In Minkowski coordinates they are
\begin{align}
    &e_{X}^{\mu}=(0,1,0,0),\hspace{5pt}e_{Y}^{\mu}=(0,0,1,0) \nonumber\\
    &e_{Z}^{\mu}=(\sinh(a\tau),0,0,\cosh(a\tau))\,.
\end{align}

The Fourier transform is then
\begin{align}\label{eq:scalardipolewightman}
    G^{(+)}_{IJ}(\Omega)=&\frac{\left(\delta_{IJ}(\Omega^{3}+a^{2}\Omega)+3\delta_{I}^{Z}\delta_{J}^{Z}a^{2}\Omega\right)}{12\pi}\left(\coth\left(\frac{\pi\Omega}{a}\right)+1\right)\,,
\end{align}
in terms of which the decoherence functional is
\begin{equation}
    \mathbb{D}=\frac{q^{2}}{2}\int \frac{d\Omega}{2\pi}\,\varepsilon^{I}(\Omega)\varepsilon^{J}(-\Omega)\left(G^{(+)}_{IJ}(\Omega)+G^{(+)}_{IJ}(-\Omega)\right)\,.
\end{equation}
As expected this is in agreement with the number of radiated photons, Eq.~\eqref{eq:DSWwightman}. For dipoles held constant for very long times, $\varepsilon^{I}(\Omega)\approx\varepsilon^{I}2\pi\delta(\Omega)$, we have the steady decoherence rate
\begin{equation}\label{eq:scalarsteadydecoh}
    \mathbb{D}\approx \frac{Tq^{2}a^{3}}{12\pi^{2}}\varepsilon^{I}\varepsilon^{J}(\delta_{IJ}+3\delta_{I}^{Z}\delta_{J}^{Z})\,. 
\end{equation}

As an aside, note that the force spectrum diverges as $\Omega^{3}$ at high frequencies.  This is the typical scaling of electric field vacuum amplitude fluctuations, and we will show, this divergence is cut off by a physical detector.

\subsection{Unruh-DeWitt Particle Detector}

A physical system with finite resolution will not be sensitive to divergent vacuum fluctuations described by Eq.~\eqref{eq:scalardipolewightman}, and we can illustrate this point by considering a simple particle detector model. This particle detector model will also give us yet another way of understanding the decoherence rate of the superposed charged particle.

Suppose now that we utilize a constant dipole $q\varepsilon^{A}$ as an Unruh-DeWitt detector for $\phi$ particles. That is, we apply a local bias field to our two-state system such that they now have an energy gap $\Delta$, and we couple two system to a scalar field via a dipolar interaction. Evolving the joint system perturbatively, we will then have a final state of 
\begin{equation}
\int d\tau 
\left[\sigma_+ e^{-i\Delta \tau} +
\sigma_- e^{+i\Delta \tau}\right]q\varepsilon^\mu(\tau) \partial_\mu \hat \phi(\bar{z}(\tau))\, |g\rangle |0\rangle,
\end{equation}
where $|g\rangle$ is the ground-state of the two-level system and $|0\rangle$ is the ground state of the field.  The probability that the system detects a particle, i.e. gets excited from its ground state, is then given by the detector response function
\begin{equation}
    \mathscr{F}(\Delta)=q^{2}\varepsilon^{I}\varepsilon^{J}\int_{-\infty}^{0}d\tau\int_{-\infty}^{\infty}d\tau'\,e^{-i\Delta(\tau-\tau')}\,G^{(+)}_{IJ}(\bar{z}(\tau),\bar{z}(\tau'))\,.
\end{equation}

In the language of the previous section, we effectively have
\begin{equation}
    \varepsilon^{I}(\Omega) = \varepsilon^{I} 2\pi\delta(\Omega-\Delta).
\end{equation}
This will lead to a photodetection rate of 
\begin{equation}
dN/dt = q^{2}\varepsilon^{I}\varepsilon^{J}G^{(+)}_{IJ}(-\Delta )=\frac{q^{2}\varepsilon^{I}\varepsilon^{J}}{6\pi}\frac{\left(a^{2}\Delta(\delta_{IJ}+3\delta_{I}^{Z}\delta_{J}^{Z})+\Delta^{3}\right)}{e^{2\pi\Delta/a}-1}\,.
\end{equation}
This features a Bose distribution with temperature $\beta^{-1}=a/2\pi$, characteristic of the Unruh effect, however it is not equal to the Planck spectrum. The photodetection rate is finite, and for systems with large energy gaps, $\Delta\gg \beta^{-1}$, the rate is exponentially suppressed. As expected on physical grounds, a photodetector will not suffer from the same divergence as a detector with infinite resolution to amplitude fluctuations.

From this expression we may also see another interpretation of the DSW decoherence rate. If we take the gap size to zero, the photodetector clicks with a rate
\begin{equation}\label{eq:zeroenergyparticledetector}
dN/dt = \frac{q^2 a^3}{12 \pi ^2}\varepsilon^{I}\varepsilon^{J}(\delta_{IJ}+3\delta_{I}^{Z}\delta_{J}^{Z})\,.
\end{equation}
This agrees exactly with the number of radiated soft photons, Eq.~\eqref{eq:charradiatedphotonnum}, as seen in the inertial frame. It also agrees with the decoherence predicted by the description of local random forces acting on the dipole, Eq.~\eqref{eq:scalarsteadydecoh}. The agreement with Eq.~\eqref{eq:zeroenergyparticledetector} suggests yet another interpretation from the perspective of the lab frame, the two branches of the dipole's wavefunction randomly mix with one another due to the absorption of soft photons from a thermally populated bath.

\begin{figure*}
\includegraphics[width=0.95\textwidth]{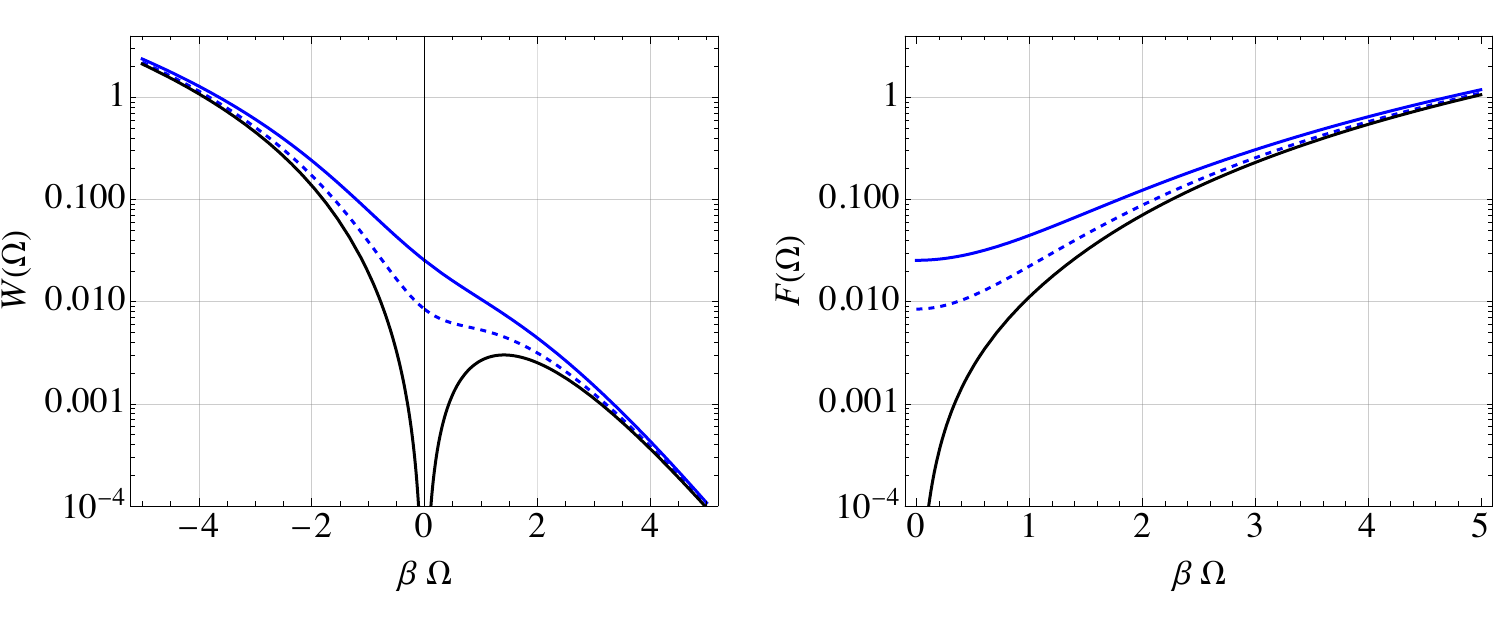}
\caption{Unruh-DeWitt versus thermal Wightman functions.  Left panel: we use the blue solid line to denote the Wightman function of a dipole along the direction of acceleration, blue dashed line for a dipole along transverse directions, and black solid line for any dipole in a thermal field (in this case, the Wightman function is orientation independent). The right panel shows the symmetrized Wightman function, which represents the force spectrum acting on the dipole. The positive-frequency portion of the left panel also indicates photon counts by an Unruh-DeWitt detector along the accelerating path.  The three curves all exponentially decay and agree with each other at high frequencies.  Photon fluxes computed by DSW agree with the Unruh-DeWitt Wightman functions.  Their disagreement with the thermal Wightman function reflects the different way an accelerating dipole couples to a scalar field, rather than the state of the scalar field.  }
\label{fig:wightman}
\end{figure*}

\subsection{Electric Dipole Decoherence}

Let us now consider the more physically relevant system of a delocalized charged particle. From the general discussion in Sec.~\ref{sec:generalformalism}, we know that for our purposes the generic two-path Wilson loop expression, Eq.~\eqref{eq:decoherencefunctionalloopintegral}, can be simplified down to a dipole approximation
\begin{equation}
    \mathbb{D}=\frac{q^{2}}{2}\int_{-\infty}^{\infty} d\tau d\tau'\,\varepsilon^{A}(\tau)\varepsilon^{J}(\tau)\, \Big\langle \{E_{A}(\tau),E_{J}(\tau')\}\Big\rangle\,,
\end{equation}
where $E^{A}(\tau)$ is the electric field in the local orthonormal frame in the lab following worldline $\bar{z}^{\mu}(\tau)$. Again, the field is taken to be in the Minkowski vacuum state.

Since the decoherence functional is gauge invariant, we may compute the above correlation function by evaluating certain derivatives of the Feynman gauge Wightman function
\begin{equation}
\left\langle 0| A_{\mu}(x)\,A_{\nu}(y) |0\right\rangle=\frac{-\eta_{\mu\nu}}{4\pi^{2}}\frac{1}{(x^{0}-y^{0}-i\epsilon)^{2}-(\vec{x}-\vec{y})^{2}}\,,
\end{equation}
to obtain
\begin{align}
    G^{(+)\,{\rm EM}}_{IJ}(\tau,\tau')&\equiv\left\langle 0| E_{I}(\tau)\,E_{J}(\tau') |0\right\rangle\nonumber \\
    &=\frac{\delta_{IJ}a^{4}}{16\pi^{2}}\left(\frac{1}{\sinh\left(\tfrac{a}{2}(\tau-\tau')\right)-i\epsilon}\right)^{4}\,.
\end{align}
In contrast with the scalar case, this is actually isotropic in the local frame. The Fourier transform is
\begin{equation}
    G^{(+)\,{\rm EM}}_{IJ}(\Omega)=\delta_{IJ}\frac{\Omega^{3}+a^{2}\Omega}{6\pi}\left(\coth\left(\frac{\pi\Omega}{a}\right)+1\right)\,,
\end{equation}
and the steady decoherence rate is
\begin{equation}\label{eq:EMsteadydecoh}
    \mathbb{D}\approx \frac{Tq^{2}|\vec \varepsilon\,|^{2}a^{3}}{6\pi^{2}}\,.
\end{equation}

Here we have an isotropic rate of decoherence regardless of which spatial direction the superposition is along.    Since $G^{(+)\,{\rm EM}}_{IJ}$ is constant as $\Omega \rightarrow 0$, we still have a decoherence level proportional to proper time $T$. In the time-domain this non-vanishing zero-frequency limit corresponds to electric field correlations which do not average out over arbitrarily long times, i.e.
\begin{equation}
    \langle 0| {\left(\int_{-\infty}^{\infty} d\tau\, E_{I}(\tau)\right) \,E_{J}(0)} |0\rangle = \frac{a^{3}}{6\pi^{2}}\,.
\end{equation}
Given the discussion around Eq.~\eqref{eq:stochasticdecoh}, it follows that the electric charge undergoing superposition feels a stochastic force which remains correlated over arbitrarily long times, never averaging out, and always causing dephasing on the quantum system.

As we contrast the EM result Eq.~\eqref{eq:EMsteadydecoh} with the scalar case Eq.~\eqref{eq:scalarsteadydecoh}, we see that for the $X$ and $Y$ directions, the EM decoherence rate is double the scalar case.   For the $Z$ direction, the scalar case is double the EM case.

More importantly, we note that the fluctuation spectrum for the local electric field is
\begin{equation}
    \frac{G^{(+)\,{\rm EM}}_{IJ}(\Omega)+G^{(+)\,{\rm EM}}_{IJ}(-\Omega)}{2}=\delta_{IJ}\frac{\Omega  \left(a^2+\Omega ^2\right) }{3 \pi }\left(\frac{1}{2}+\frac{1}{e^{2\pi\Omega/a}-1}\right),
\end{equation}
which differs crucially from the naive Planck spectrum for thermal fluctuations in an inertial lab,
\begin{equation}
    S_{IJ}(\Omega)=\delta_{IJ}\frac{\Omega ^3}{3 \pi }\left(\frac{1}{2}+\frac{1}{e^{\beta\Omega}-1}\right)\,,
\end{equation}
by the addition of a linear term multiplying the Bose factor. This difference is crucial to explaining why the decoherence observed by DSW is not \emph{just} explained by thermal radiation bombarding the quantum state. 

The difference between the accelerated and thermal case can be seen in the force spectrum, and can also be seen clearly at the level of the Wightman function. A thermal correlator of electric fields in an inertial rest frame is 
\begin{equation}
    \int dt\, e^{i\Omega t}\textrm{Tr}(e^{-\beta H}E_{I}(t)E_{J}(0))=\delta_{IJ}\frac{\Omega^{3}}{6\pi}\left(\coth\left(\frac{\beta\Omega}{2}\right)+1\right)\,.
\end{equation}
We show the $z$-dipole and $x$-dipole Unruh-DeWitt Wightman functions together with the thermal one at the same temperature, as well as the $F$ functions, in Figure~\ref{fig:wightman}. 

We've seen that the steady rate of decoherence is not trivially explained by considering the same experiment in an inertial thermal bath, however, in the following section we will see that the existence of the thermal bath of Unruh photons is the only relevant \emph{quantum mechanical} effect. The slight modification of the Wightman function will be understood from classical considerations alone, and is in fact already apparent from the familiar Abraham-Lorentz-Dirac radiation reaction force.

\section{General Discussions on Decoherence Rate and Bath Temperature}
\label{sec:relationtoFDT}

This is a well-documented feature in the literature, as discussed by Ref.~\cite{lima2019probing} and references therein.   However, we provide a more pedagogical treatment here in our text.  

\subsection{The KMS State and the End State of Thermalization}

Let us first show that while the end state of our quantum system is a thermal one,
\begin{equation}
    p_n =\frac{1}{Z} e^{-\frac{E_n}{k_B T}}=
    \frac{1}{Z} e^{-\frac{2\pi E_n c}{\hbar a}}
    \,,\quad \beta=1\,,
\end{equation}
its dynamical path toward that thermal state will be different.  Let us first explain why the end state of the accelerating quantum system should be thermal with temperature $\beta = 2\pi\hbar a/(k_B c)$.  For any field operators $ A$ and $B$, for example, electric and magnetic fields or their derivatives in the rest frame of the constantly accelerating observer, we have the so-called Kubo-Martin-Schwinger (KMS) property: 
\begin{equation}
    \langle A (\tau_1)  B  (\tau_2)\rangle  =
        \langle B(\tau_2)  A  (\tau_1+ 2\pi i n /a)\rangle  \,,\quad n\in \mathbb{Z}\,.
\end{equation}
We can explicitly see this in calculations in Secs.~\ref{sec:decoherencefunctional}, because all dependence on $\tau$ are via $e^{\pm  a\tau}$.  Typically, we expect the system to gain the same periodicity as the bath --- as is true for real-valued periods.  If this is indeed the case, then for a finite-dimensional system, we can consider the operator
\begin{equation}
\hat P_{mn} = |m\rangle \langle n |
\end{equation} 
with $|m\rangle$ and $|n\rangle$ the energy eigenstates of the system. For the Heisenberg operators of $\hat P_{mn}$, we expect
\begin{equation}
\langle \hat P_{mn} (t) \hat P_{nm} (t) \rangle =
\langle \hat P_{nm} (t) \hat P_{mn} (t+2\pi i a)  \rangle
\end{equation}
This leads to 
\begin{equation}
    \rho_{mm} = e^{2\pi a (E_n-E_m)/\hbar}\rho_{nn}
\end{equation}
which indeed corresponds to a thermal state with
\begin{equation}
    \beta = 2\pi a/\hbar\,.
\end{equation}
Having the same temperature only means the same end state of thermalization --- it does not guarantee the same decoherence rate, as the time it takes to reach the end state can be different.

\subsection{Fluctuation-Dissipation Theorem and the Process Toward Thermalization}

To illustrate the possible difference in decoherence rates, let us consider a dipole antenna that is part of an oscillator, and  use that antenna to measure the temperature of the field.  Suppose we have
\begin{equation}
\hat H = \hat H_A - \hat x \hat F + \hat H_{\rm field} \,,
\end{equation} 
where the two-time correlation function of $\hat F$, in absence of coupling, is given by 
\begin{equation}
\langle     \hat F^{(0)}(\tau_1) \hat F^{(0)}(\tau_2) \rangle=  W(\tau_1,\tau_2)\,.
\end{equation}
We can further define 
\begin{equation}
\langle     \hat F^{(0)}(\tau_1) \hat F^{(0)}(\tau_2) \rangle_{\rm sym} =  \frac{W(\tau_1,\tau_2)+W(\tau_2,\tau_1)}{2}
\end{equation}
and 
\begin{equation}
[ \hat F^{(0)}(\tau_1) , \hat F^{(0)}(\tau_2) ] =    \frac{W(\tau_1,\tau_2)-W(\tau_2,\tau_1)}{2}
\end{equation}
For a steady state system, we have $W(\tau_1,\tau_2) = W(\tau_1-\tau_2)$ and 
\begin{equation}
\label{SFW}
S_F =\frac{W(\Omega) + W(-\Omega)}{2}\,.
\end{equation}
We also define response functions of $\hat x$ and $\hat F$, 
\begin{equation}
    \chi_A = i\int_0^{+\infty} d\tau e^{i\Omega\tau }[\hat A^{(0)}(\tau), \hat A^{(0)}(0)]\,.
\end{equation}
One can show that 
\begin{equation}
    \mathrm{Im}[\chi_F(\Omega)] =\frac{W(\Omega) - W^*(-\Omega)}{2}\,.
\end{equation}
In this way, we have related the symmetric and anti-symmetric Wightman function (in the time domain) to the force spectrum given by the bath and the response function of the bath to the system.  We will now show that this is also related to the damping the system experiences once coupled to the bath.

Now that the oscillator and the bath are coupled, we have 
\begin{align}
\hat x^{(1)} &= \hat x^{(0)} +\chi_x[ \hat F^{(1)}+ G] \,,\\
\hat F^{(1)} &= \hat F^{(0)} +\chi_F \hat x^{(1)} \,.   
\end{align}
From this we can obtain
\begin{equation}
    \hat x^{(1)} = \left[ \frac{1}{\chi_x} +\chi_F\right]^{-1}\left[ G +F^{(0)}\right]\,.
\end{equation}
In this way, we find that the inverse response of the oscillator changes by $\chi_F$, or 
\begin{equation}
\left[    \frac{1}{\chi_x}\right]^{(1)} =
\left[    \frac{1}{\chi_x}\right]^{(0)} +\chi_F
\end{equation}
If the oscillator originally has no damping to any other bath, we will then have
\begin{equation}
    \mathrm{Im} \chi_x = \frac{W(\Omega)-W^*(-\Omega) }{2}
\end{equation}
If we were to apply the fluctuation-dissipation theorem, for a bath inverse temperature $\beta$, we will have
\begin{equation}
\label{eq:SFDT}
    S_F^{\rm FDT} = \coth\frac{\beta\Omega}{2}\mathrm{Im}\chi_x 
\end{equation}
In order to reconcile Eq.~\eqref{eq:SFDT} and Eq.~\eqref{SFW}, we need to impose 
\begin{equation}
W(\Omega)+ W(-\Omega) = \coth \beta\Omega \left[W(\Omega)- W(-\Omega)\right]
\end{equation}

We can now apply the above formalism to the most interesting electromagnetic case.  For a charge $q$, we have 
\begin{equation}
    W^{\textrm{EM}}_{IJ}(\Omega) = \delta_{IJ}\frac{q^2 a^2}{6\pi}\Omega\left[1+\left(\frac{\Omega}{a}\right)^2\right] \left(\coth\frac{\pi\Omega}{a}+1\right)
\end{equation}
From here, we obtain 
\begin{equation}
    \mathrm{Im}[\chi^{\textrm{EM}}_{F\,AB}(\Omega)]=\frac{q^2(\Omega^3+a^2\Omega)}{6\pi}\,.
\end{equation}
This reduces to the vacuum case when $a\rightarrow 0$, recovering the Abraham-Lorentz damping. Similarly enhanced low frequency damping in accelerated frames has been discussed previously outside the context of decoherence (see e.g.~\cite{PhysRevD.73.124018, Iso:2010yq, Oshita:2015xaa, PhysRevD.95.023512}).

\subsection{General Relationships Between Decoherence and Friction}
The preceding computations  invite a more general analysis. A natural question to address the question is, under what circumstances will one find steady decoherence. In Sec.~\ref{sec:thermaldecoherencereview} we illustrated the monopolar coupling example, and in the other sections we went through a number of computations for dipolar couplings.  In both cases we found that experiments done in the accelerated lab underwent decoherence, however only for the monopolar coupling could this be entirely explained via a thermal bath of Unruh radiation. To understand this, we return to the general formula for decoherence, Eq.~\eqref{eq:generalDecoherence}.

Consider the simplest setting which parallels the cases we're interested in, that of bilinear coupling: $Q(q)=q$, $\Phi(\phi)=\phi$. Let $q^{+}(\tau)=\tfrac{1}{2}\varepsilon(\tau)$, $q^{-}(\tau)=-\tfrac{1}{2}\varepsilon(\tau)$, where $\varepsilon(\tau)$ vanishes at early and late times. We'll also assume that $\langle\phi\rangle$=0, and that $\phi$ is a free-field in the sense that $[\phi(\tau),\phi(\tau')]$ is a $c$-number. The decoherence functional then has the form
\begin{equation}
    \mathbb{D}=\frac{1}{2}\int_{-\infty}^{\infty}d\tau\int_{-\infty}^{\infty}d\tau'\,\varepsilon(\tau)\varepsilon(\tau') \bigg\langle\{\phi(\tau),\phi(\tau')\}\bigg\rangle\, .
\end{equation}
Let us focus on the case that $\phi(\tau)$ is a finite temperature quantum system.  We may then use the fluctuation-dissipation theorem to relate the spectrum of the fluctuations to the response function,
\begin{equation}
    \int d\tau e^{i\omega\tau}\Big\langle\{\phi(\tau),\phi(0)\}\Big\rangle_{\beta}=2\coth\left(\frac{\beta \omega}{2}\right)\textrm{Im}[\chi_{\phi}(\omega)]\,,
\end{equation}
where
\begin{equation}
    \textrm{Im}[\chi_{\phi}(\omega)]=\frac{1}{2}\int d\tau e^{i\omega \tau} \langle 0 | \phi(\tau)\phi(0)-\phi(0)\phi(\tau) |0\rangle\,.
\end{equation}

In the long time limit, with $\varepsilon$ constant, we then find the leading contribution to the decoherence of the form $\frac{1}{2}\mathbb{D}=\Gamma_{\beta} T$, with rate
\begin{equation}\label{eq:generalsteadydecoh}
\Gamma_{\beta} = \frac{\varepsilon^{2}}{\beta} \lim_{\omega\rightarrow 0}\left(\frac{\textrm{Im}[\chi_{\phi}(\omega)]}{\omega}\right)\,.
\end{equation}
We see that the stationary rate of decoherence is determined by the linear part of the spectral density.

While the linear part of the spectral density plays a central role in determining the decoherence rate, it also arises in a more familiar classical physics context, namely, friction. Taking the bilinear $q\phi$ interaction, we can ask how much energy is dissipated into $\phi$ if we fix a particular classical trajectory for $q(\tau)$. Classical linear response theory then gives the rate of dissipation
\begin{equation}
    \dot{E}=q(\tau)\frac{d}{d\tau}\int d\tau'\,\chi_{\phi}(\tau-\tau')q(\tau')\,.
\end{equation}
If the system is driven periodically such that its velocity is given by $\dot{q}=v \cos(\omega \tau)$, then it's straightforward to evaluate the integral and find the cycle-averaged dissipation rate
\begin{equation}
    \dot{\bar{E}}_\omega=2v^{2}\frac{\textrm{Im}[\chi_{\phi}(\omega)]}{\omega}\,.
\end{equation}
The zero frequency part of this, relevant to constant motion $q(\tau)=v\tau$, is related to our decoherence rate. The energy dissipated via constant motion with velocity $v$ is directly related to the friction coefficient, $\gamma_{1}$, defined by the Ohmic friction force $F=-\gamma_{1} \dot{q}$,
\begin{equation}
     \lim_{\omega\rightarrow 0}\dot{\bar{E}}_\omega=2v^{2}\gamma_{1}\,,
\end{equation}
which immediately identifies the linear part of the spectral density with the Ohmic friction coefficient $\gamma_{1}$.

Taking this together, we have a simple universal relation between the decoherence rate and friction force induced by a thermal bath,
\begin{equation}\label{eq:decoherencefrictionrelation}
    \Gamma_{\beta}=\frac{\varepsilon^{2}}{\beta}\gamma_{1}\,.
\end{equation}
Said again, if an environment induces a simple Ohmic friction force on a system (proportional to the first derivative of the system) then when that environment is at a finite temperature it will invariably cause a constant rate of decoherence in the system. This generic behaviour is well known in the field of open quantum systems~\cite{Moore:1984bk}.

This argument can be made more quantitative if one starts from the influence functional, Eq.~\eqref{eq:influencefunctional}, and derives an effective Langevin equation, as in~\cite{Schmid:1982qcl}. Disregarding the decoherence functional, the remaining part of the influence functional is given by the off-diagonal effective action
\begin{align}
    \tilde{S}[q^{+},q^{-}]&=\frac{i}{2}\int^{t}_{-\infty}d\tau\int^{\tau}_{-\infty}d\tau'\,\Big\langle[\phi(\tau),\phi(\tau')]\Big\rangle \nonumber \\
    &\times\Big(q^{+}(\tau)-q^{-}(\tau)\Big)\Big(q^{+}(\tau')+q^{-}(\tau')\Big)\,.
\end{align}
The classical motion of the central system is described by the evolution of its expectation value, $\langle q(t)\rangle$, and in the in-in path-integral this is equivalently computed by either $\langle q^{+}\rangle$ or $\langle q^{-}\rangle$. Hence, a natural change of variables is to $X=\tfrac{1}{2}(q^{+}+q^{-})$ and $\xi=q^{+}-q^{-}$. To find the effective classical dynamics for $q$ one expands the effective action to leading order in $\xi$, 
\begin{align}
&S^{\rm eff}=S_{q}[q^{+}]-S_{q}[q^{-}]+\tilde{S}[q^{+},q^{-}] \nonumber\\
&\approx \int d\tau\,\xi(\tau)\left[ \frac{\delta}{\delta X(\tau)} S_{q}[X]-\int d\tau'\chi_{\phi}(\tau-\tau')X(\tau')\right]\,,
\end{align}
where the response function is given by the Kubo formula
\begin{equation}
    \chi_{\phi}(\tau-\tau')=-i\theta(\tau-\tau')\,\big\langle[\phi(\tau),\phi(\tau')]\big\rangle\,.
\end{equation}

The classical equation of motion for $X$ follows from a $\xi$ variation, 
\begin{equation}
  \frac{\delta}{\delta X(\tau)} S_{q}[X]+\tilde{F}(\tau)=\eta(\tau)\,,
\end{equation}
the first term is the equation of motion in the absence of the bath, $\eta(\tau)$ is a stochastic forcing term with noise spectrum determined by the correlation function $\langle\{\phi(s),\phi(s')\}\rangle$ appearing in the decoherence functional, and the environment mediated deterministic forces are
\begin{equation}
    \tilde{F}(\tau)=-\int d\tau'\chi_{\phi}(\tau-\tau')X(\tau')\,.
\end{equation}
Provided that the spectral density of $\phi$ is an analytic function in frequency space over the range of energies for which the effective theory is defined, we can write a low energy expansion
\begin{equation}\label{eq:generalimchi}
    \textrm{Im}[\chi_{\phi}(\omega)]=\gamma_{1}\omega +\gamma_{3}\omega^{3}+\cdots\,.
\end{equation}

In general the bath response function may also generate conservative forces, however the contributions coming from the odd part of the spectral density will lead to dissipative contributions to $\tilde{F}$\,, and for a dissipative response parameterized as Eq.~\eqref{eq:generalimchi}, one finds dissipative forces in the classical equation of motion\footnote{Generally the odd part of the spectral density also renormalizes couplings in conservative sector, and we have omitted the details of this here.} for $x$,
\begin{equation}
    \tilde{F}_{\rm diss}(\tau)=-\gamma_{1}\dot{X}(\tau)+\gamma_{3}\dddot{X}(\tau)+\cdots\,.
\end{equation}

Some relevant examples of this computation include: a particle with charge $q$ and generic worldline $X^{\mu}(\tau)$~\cite{Galley:2006gs}, and a dipole $q\varepsilon^{I}$ anchored to a  uniformly accelerated worldline $\bar{z}^{\mu}(\tau)$, for which the above computation yields, respectively,
\begin{equation}\label{eq:ALD1}
    \tilde{F}^{\mu}=\frac{q^{2}}{6\pi}\Big(\eta^{\mu\nu}+\dot{X}^{\mu}\dot{X}^{\nu}\Big)\dddot{X}_{\nu}\,,
\end{equation}
and
\begin{equation}\label{eq:dipoleradreact}
    \tilde{F}^{I}=-\frac{q^{2}a^{2}}{6\pi}\dot{\varepsilon}^{I}+\frac{q^{2}}{6\pi}\dddot{\varepsilon}^{I}\,,
\end{equation}
which are the relativistic Abraham-Lorentz-Dirac force (i.e. the familiar third derivative term suitably projected into the local orthonormal frame), and an analog formula for a dipole. By noting that $\tfrac{d^{2}}{d\tau^{2}}(\dot{X}^{\mu}\dot{X}_{\mu})=0$, the ALD formula can actually be equivalently rewritten as 
\begin{equation}\label{eq:ALD2}
    \tilde{F}^{\mu}=\frac{q^{2}}{6\pi}\left(\dddot{X}^{\mu}-\dot{X}^{\mu}\Big(\ddot{X}^{\nu}\ddot{X}_{\nu}\Big)\right)\,.
\end{equation}
From this form it is more obvious that for $X^{\mu}=\bar{z}^{\mu}(\tau)+\varepsilon^{I}e^{\mu}_{I}$, with $a\varepsilon^{I} \ll 1$ oriented in the $X,Y$ directions, the ALD formula is equal to the dipole force formula.

In general though, these formulae are not identical. The ALD force Eq.~\eqref{eq:ALD2} is expressed in spacetime indices, whereas our derived force Eq.~\eqref{eq:dipoleradreact} is expressed in the local orthonormal frame. This difference is immaterial for transverse displacements since $\partial_{\tau}e^{\mu}_{X}=\partial_{\tau}e^{\mu}_{Y}=0$, and the formulae do agree. For $\varepsilon^{\mu}$ with components tangential to the acceleration non-inertial frame effects arise which lead to a disagreement between the formulae. This disagreement is no issue though, as Eq.~\eqref{eq:dipoleradreact} is the force felt by an electric-\textit{dipole} in an accelerated frame whereas Eq.~\eqref{eq:ALD2} is the force felt by an electric-\textit{monopole} which has small displacement. These two notions agree for transverse displacements but they simply disagree for longitudinal displacements. 

Even without a careful computation of the dipolar radiation reaction force one could take the ALD force and anticipate the appearance of an Ohmic friction term.  To do so, simply omit the third-derivative ``Schott-term'' in Eq.~\eqref{eq:ALD2}, and focus only on the ``radiation'' term. This term then maps into the local orthonormal frame cleanly, and for all components of $\varepsilon$ one indeed matches the first-derivative term in Eq.~\eqref{eq:dipoleradreact}, as expected.

Returning now to our main point, we can summarize the above discussion as follows. If a system is linearly coupled to a thermal bath it will experience an irreducible constant rate of decoherence  given by Eq.~\eqref{eq:decoherencefrictionrelation}, where $\varepsilon^{2}$ characterizes the
`size' of the superposition, $\beta^{-1}$ is the bath temperature, and $\gamma_{1}$ is the coefficient of the Ohmic friction force which the bath induces in the classical equation of motion for the system. For the uniformly accelerated dipole, the Ohmic friction coefficient is $\gamma_{1}=q^{2}a^{2}/6\pi$, which leads to a thermal decoherence rate,
\begin{equation}
    \Gamma_{\beta}=\frac{\varepsilon^{2}}{\beta}\frac{q^{2}a^{2}}{6\pi}\overset{\rm Unruh}{\rightarrow} \frac{q^{2}\varepsilon^{2}a^{3}}{12\pi}\,,
\end{equation}
which exactly matches that computed by DSW when $\beta^{-1}$ is taken to be the Unruh temperature.

\section{Conclusions}
\label{sec:conclusions}

In this paper, we proved explicitly the decoherence effect found by DSW is the same as what one would obtain by going through a Unruh-DeWitt type calculation based on the local field fluctuations.  This equivalence is connected to the unitarity of quantum mechanics: decoherence caused by incoming field fluctuations disturbing the quantum system is related to information of the system flowing into the outgoing field. 

We verified the equivalence through explicit calculations considering coupling to both scalar and to electromagnetic fields, with the system undergoing superposition being split in arbitrary directions.  In all such situations, the system under constant acceleration had a constant decoherence rate proportional to $\sim q^2  \varepsilon^2 a^3$.  

Such a constant decoherence rate was, at first sight,  in conflict with thermal decoherence, because a two-path experiment along an inertial central trajectory does not have a constant decoherence rate when the ambient radiation field has a constant temperature.   We reconcile this conflict by noting that as a quantum system is undergoing acceleration, its coupling to the scalar or EM field causes an additional radiation damping at low frequencies, which restores the constant decoherence rate.  More specifically, in the EM case, the non-relativistic Abraham-Lorenz damping force is proportional to $\sim q^2 \dddot x$, while in an accelerating frame there is an additional damping force proportional to $\sim q^2a^2 \dot x$. 

More generally, as we consider a steady-state Wightman function $W$, the dephasing force (fluctuation) has a spectrum related to the symmetric part of $W$,
\begin{equation}
    S_F =\frac{W(\Omega)+W^*(-\Omega)}{2}\,.
\end{equation}
with decoherence rate given by 
\begin{equation}
     \Gamma = S_F(0)\,.
\end{equation}
On the other hand, the damping (dissipation) of the system is given by 
\begin{equation}
    \mathrm{Im}\chi_F =\frac{W(\Omega)-W^*(-\Omega)}{2}\,.
\end{equation}
The two are related via
\begin{equation}
    \frac{W(\Omega)+W^*(-\Omega)}{W(\Omega)-W^*(-\Omega)} =\coth\frac{\beta\Omega}{2}
\end{equation}
which is guaranteed by the KMS condition. The decoherence rate in a thermal bath is then
\begin{equation}
    \Gamma = \frac{2}{\beta}\lim_{\Omega\rightarrow 0}\frac{\mathrm{Im}\chi_F(\Omega)}{\Omega}\,.
\end{equation}

Even though the Danielson-Satishchandran-Wald decoherence ended up ``equivalent'' to the Unruh-DeWitt decoherence, their insight about a steady stream of soft photons escaping across the Killing horizon is still the most efficient and elegant way to understand the  physical mechanism for this decoherence effect.   Now that we understand this equivalence, we can see that the DSW mechanism not only underlies the constant decoherence rate, but more importantly also underlies the elevated level of radiation reaction for an accelerating quantum system.

Let us now summarize the logic of how an accelerated observer, Alice, would explain the observed decoherence of her experiment. In the accelerated lab her classical electronics function slightly differently. In particular, the radiative losses from her antenna (with dipole moment $q\varepsilon^{I}$) have a form described at low frequencies by an Ohmic friction force $F^{I}=-\frac{q^2}{6\pi}a^{2}\dot{\varepsilon}^{I}$. Alice can then deduce that at low frequencies, the local electric-field has a linear spectral density $\textrm{Im}[\chi_{IJ}(\Omega)]=\delta_{IJ}\frac{q^{2}a^{2}}{6\pi}\Omega$. 

She wants to perform an experiment in which she places the dipole into a superposition state and holds it for a very long time, and she naively anticipates that a vacuum-state electromagnetic field will cause no issue because the vacuum spectral density vanishes at low frequencies. The problem for Alice, though, is that the Unruh effect forces her accelerated lab to feel a finite temperature $\beta^{-1}=a/2\pi$. As a result, the electromagnetic field is instead in a thermal state and the divergent thermal population of low energy modes $n_{B}(\Omega)+\tfrac{1}{2}\approx (\beta\Omega)^{-1}$ compensates the linearly vanishing spectral density and leads to a flat spectrum of low energy thermal electric field fluctuations. Alice's superposed dipole experiences thermal electric field fluctuations correlated over all time scales, and decoheres exponentially with a constant rate.

The explicit calculations is this paper involved only Rindler horizons, however the extension to discussions of black hole horizons is clear. The drastic Bose enhancement of the Wightman function at low frequency will occur, provided that the black hole is subextremal and that the state of the quantum fields propagating on the black hole spacetime is the Hartle-Hawking state or the more physical Unruh vacuum. Given this enhancement, all that is required to find constant decoherence is that the response function for the fields propagating on the black hole spacetime has a leading Ohmic behaviour, ie. absorption linear in frequency. This behaviour has been demonstrated, in detail, in the effective field theory of black hole horizons ~\cite{Goldberger:2019sya,Goldberger:2020geb}--indeed the authors already discussed the low frequency enhancement of the Wightman function in thermal states that we've emphasized here. The authors of that work noted that the low frequency enhancement would only be observable in Planck suppressed corrections to classical processes, since the only correlator measurable in a classical process is the \emph{anti-}symmetric part of the Wightman function, ie. retarded Green's function. The decoherence suffered by Alice's probe is a quantum mechanical probe of the state of the fields, and so can be sensitive to ``state-dependent'' correlators. Studying decoherence is perhaps the simplest setting in which one is indeed sensitive to the aforementioned low frequency enhancement.

 In this work we have studied the decoherence of stationary superpositions of charged particles (coupled to either scalar or electromagnetic fields). As shown by DSW, for an arbitrarily slowly prepared superposition, a finite amount of decoherence is inevitable for superpositions in the frame of a Rindler observer, or at rest in Schwarzschild or de Sitter spacetimes---and moreover the off-diagonal density matrix elements decay exponentially with constant rate. We verify these results, in the specific case of Rindler observers, via an equivalent but complementary \emph{local} description of the decoherence. The finite Unruh temperature felt in the accelerating lab plays a central role, as it leads to an drastic enhancement of locally measured low frequency electric field fluctuations. These fluctuations induce a stochastic force on the delocalized charged particle for which the auto-correlations do not self-average over arbitrarily long time periods---resulting in inevitable dephasing. 
 
 The approach set-up in this paper,  utilizing the Feynman-Vernon influence functional and worldline-localized radiation effective field theory, can be readily be generalized to discuss: gravitational radiation, radiation from higher multipole moments, and observers in the presence of generic Killing horizons, however this will be left for future work. 
 
{\bf Note:} Nearly coincidentally with the submission of this work, \cite{Biggs:2024dgp} appeared which has some overlap with the content of this paper.
 
\section{Acknowledgements}
The authors thank Gautam Satishchandran, Daine Danielson, and Hongji Wei for useful discussions.  Research of Y.C., A.D., and J.W.G. is supported by the Simons Foundation (Award No.\ 568762), and NSF Grant PHY-2011968. J.W.G. is additionally supported by a fellowship at the Walter Burke Institute for Theoretical Physics, a Presidential Postdoctoral Fellowship, and the
DOE under award number DE-SC001163.

\bibliography{references}

\clearpage
\newpage
\appendix
\begin{widetext}
\section{Photon Number in the Characteristic Formulation} 
\label{sec:jacobian}
For completeness, I will discuss why the integral used in the characteristic formulation is the correct one for evaluating photon fluxes.  Suppose we decompose
\begin{equation}
\phi(x^\mu) = \int \frac{d^3\mathbf{k}}{(2\pi)^3\sqrt{2\omega_{\mathbf{k}}}} 
\left[
e^{-i\omega t + i k_x x + i k_y y + i k_z z}  \phi_{k_x,k_y,k_z} +
e^{+i\omega t - i k_x x  - i k_y y -  i k_z z}  \phi^*_{k_x,k_y,k_z}
\right]\,,
\end{equation} 
we can obtain $\phi_{k_x,k_y,k_z}$ from the Fourier transform
\begin{equation}
\Phi_{\Omega,\tilde k_x,\tilde  k_y} = 
\int  d\xi dx dy  
\left[  e^{i\Omega \xi -i \tilde k_x x - \tilde k_y y } \phi(\xi,x,y,\xi)
\right]
\end{equation} 
We obtain
\begin{equation}
\Phi_{\Omega,\tilde k_x,\tilde  k_y}
 =\int{dk_z}
 \frac{ \delta\Big(\Omega-\sqrt{\tilde k_x^2+\tilde k_y^2+k_z^2}+k_z\Big)
}{\sqrt{2}(\tilde k_x^2+\tilde k_y^2+k_z^2)^{1/4}}\phi_{\tilde k_x,\tilde k_y,k_z}
 +\int{dk_z} 
 \frac{ \delta\Big(\Omega+\sqrt{\tilde k_x^2+\tilde k_y^2+k_z^2}-k_z\Big)
}{\sqrt{2}(\tilde k_x^2+\tilde k_y^2+k_z^2)^{1/4}}
 \phi^*_{-\tilde k_x,-\tilde k_y,k_z}
\end{equation}
 For $\Omega \ge 0  $, we will only have the first branch.   
 \begin{equation}
 \delta\Big(\Omega-\sqrt{\tilde k_x^2+\tilde k_y^2+k_z^2}+k_z\Big)
=\frac{\tilde k_x^2+\tilde k_y^2 +\Omega^2}{2\Omega^2}\delta\left( k_z - \frac{\tilde k_x^2+\tilde k_y^2-\Omega^2}{2\Omega} \right)
 \end{equation} 
 This leads to 
 \begin{equation}
\Phi_{\Omega,\tilde k_x,\tilde  k_y} = 
\frac{1}{2}\frac{\sqrt{  \tilde k_x^2+\tilde k_y^2+\Omega^2}}{\Omega^{3/2}}
\phi_{\tilde k_x,\tilde k_y, \frac{\tilde k_x^2+\tilde k_y^2-\Omega^2}{2\Omega} }
 \end{equation}
 We note that as long as $\tilde k_x^2+\tilde k_y^2$ does not vanish, the $\tilde k_z$ value is able to span all real numbers.  Here we still remember that $\Omega$ is chosen to be positive.  In this way, 
\begin{equation}
[\Omega,\tilde k_x, \tilde k_y] \in R^+\times R\times R
\end{equation}
is a new coordinate system for the $(k_x,k_y,k_z)$ space.  Note that
\begin{equation}
\mathcal{N} = \int \frac{d  k_x d  k_y d k_z}{(2\pi)^3} |\phi_{ k_x, k_y,k_z}^2|
\end{equation} 
We can rewrite this integral in terms of $(\Omega,\tilde k_x,\tilde k_y)$, with 
\begin{equation}
\tilde k_x = k_x, \;\tilde k_y = k_y,\;\ \Omega = \sqrt{k_x^2 + k_y^2+k_z^2} - k_z\,.
\end{equation}
where 
\begin{equation}
\left|\frac{\partial (\tilde k_x,\tilde k_y,\Omega)}{\partial( k_x,k_y,k_z)}\right| =\frac{ \sqrt{k_x^2 + k_y^2+k_z^2} -k_z}{ \sqrt{k_x^2 + k_y^2+k_z^2} }
\end{equation}
In this way, we have
\begin{equation}
\mathcal{N} = \iint \frac{d\tilde k_x d\tilde k_y }{(2\pi)^2}\int_0^{+\infty}  \frac{d\Omega}{2\pi} 2\Omega|\Phi^2_{\Omega,\tilde k_x,\tilde k_y}|
\end{equation}

\section{Influence Functional Approach to Decoherence} 
\label{sec:influencefunctional}

To keep this article self-contained, in this Appendix we will review a derivation of the influence functional description of an open quantum system. We'll consider a \textit{central system} with degrees of freedom $q_{A}$, coupled to an environmental \textit{bath} system, with degrees of freedom $\phi_{j}$. Here $A$ and $j$ are generic indices which may include spacetime tensor indices, spatial coordinates, and other internal labels. 

We'll assume an initially uncorrelated density matrix in the asymptotic past and a joint unitary evolution of the mutually interacting systems. The reduced density matrix describing all observations made on the central system at time $t$ is then
\begin{equation}
    \bar{\rho}_{q}=\mathrm{Tr}_{\phi}\big\{U(t,\infty)\rho_{q}\rho_{\phi}U^{\dagger}(t,-\infty)\big\}.
\end{equation}
We'll describe the evolution operator $U$ via a path-integral, and we'll write the interactions between central system and bath in the generic form
\begin{equation}
    S_{\rm int}[q,\phi]=\int d\tau\,Q^{a}(q)\Phi_a(\phi),
\end{equation}
where $a$ are generic indices to be summed over, and $Q, \Phi$ are functions of the variables $q,\psi$ and their derivatives. For the cases of interest here we can assume that $Q,\Phi$ are commuting numbers, rather than Grassman variables.

Elements of the reduced density matrix then have the \textit{in-in} (or Schwinger-Keldysh~\cite{Schwinger:1960qe,Keldysh:1964ud}) path-integral expression
\begin{equation}
    \bar{\rho}_{q}(q,q')=\int^{(q,q')}_{\rho_{q}}\mathcal{D}q^{+}\mathcal{D}q^{-}\,e^{iS_{q}[q^+]-iS_{q}[q^-]}\mathcal{F}[Q^{+},Q^{-}],
\end{equation}
where the notation on the path-integral indicates that the initial state is to be $\rho_{q}$, and that at time $t$, the variable $q^+$ is set equal to $q$ while the variable $q^-$ is set equal to $q'$. The action $S_q[q]$ describes the central system in the absence of coupling to the bath. The relative sign between actions comes, of course, from the $(+)$ variables describing $U$ and the $(-)$ variables describing $U^{\dagger}$.

The influence functional is defined as
\begin{equation}
    \mathcal{F}[Q^{+},Q^{-}]=\oint_{\rho_{\phi}}\mathcal{D}\phi^{+}\mathcal{D}\phi^{-}\,e^{iS_{\phi}[\phi^+]-iS_{\phi}[\phi^-]}\,e^{iS_{\rm int}[q^+,\phi^+]-iS_{\rm int}[q^-,\phi^-]} \,,
\end{equation}
where the integral notation here indicates that the fixed boundary conditions $(\phi,\phi')$ at time $t$ are to be set equal to one another and summed over, i.e. traced. For the purpose of computing the influence functional, $Q^\pm$ serve as fixed external sources. The influence functional may then be written in the condensed form
\begin{equation}
    \mathcal{F}[Q^{+},Q^{-}]=\langle e^{i\int d\tau\,\left(Q^{a}(q^+)\Phi_a(\phi^+)-Q^{a}(q^-)\Phi_a(\phi^-)\right)} \rangle,
\end{equation}
where the expectation values are taken in the initial bath state $\rho_{\phi}$ and are defined via the above path-integral.

We will now specialize to the case where the \textit{connected} $n$-point correlation functions, i.e. cumulants, of the $\Phi$ operators are suppressed for $n>2$. Common examples include: weak coupling between the system and bath, $\Phi$ being a linear function with $\phi_{j}$ being a collection of weakly coupled or free oscillators, or the $\Phi$ being good mean-field variables in an interacting system with a factorizing large-$N$ limit (e.g. ``single-trace'' operators ). The standard cumulant expansion then gives 
\begin{align}
\mathcal{F}[Q^{+},Q^{-}]=\exp\Bigg[&\,i\int^{t}_{-\infty} d\tau\,\Big\langle Q^{+\,a}\Phi^{+}_{a}-Q^{-\,a}\Phi^{-}_{a}\Big\rangle \nonumber \\
&-\frac{1}{2}\int^{t}_{-\infty} d\tau\int^{t}_{-\infty} d\tau'\,\bigg\langle\Big(Q^{+\,a}(\tau)\Phi^{+}_{a}(\tau)-Q^{-\,a}(\tau)\Phi^{-}_{a}(\tau)\Big)\Big(Q^{+\,b}(\tau')\Phi^{+}_{b}(\tau')-Q^{-\,b}(\tau')\Phi^{-}_{b}(\tau')\Big)\bigg\rangle_{\rm conn.}\Bigg]+\cdots\,.
\end{align}

The standard in-out path-integral generates time-ordered correlation functions, however operators in the in-in path-integral carry the additional $(\pm)$ label indicating on which section of the closed time-contour (from $-\infty$ to $t$ and back again) the operator is living. The generalization of the usual time-ordering rule is rather simple: for a string of operators with various $(\pm)$ labels one should order all of the $(+)$ operators to the right of all the $(-)$ operators, time-order the string of $(+)$ operators amongst themselves, and anti-time-order the string of $(-)$ operators. With this understanding we can separate the various correlation functions which appear into combinations which manifest the real (anti-commutator) and imaginary (commutator) parts,
\begin{align}\label{eq:influencefunctional}
\mathcal{F}[Q^{+},Q^{-}]=\exp\Bigg[&\,i\int^{t}_{-\infty} d\tau\,\Big\langle \Phi_{a}\Big\rangle\bigg(Q^{+\,a}-Q^{-\,a}\bigg)-\frac{1}{2}\int^{t}_{-\infty}d\tau\int^{\tau}_{-\infty}d\tau'\,\Big(Q^{+\,a}(\tau)-Q^{-\,a}(\tau)\Big)\Big(Q^{+\,b}(\tau')+Q^{-\,b}(\tau')\Big)\Big\langle[\Phi_{a}(\tau),\Phi_{b}(\tau')]\bigg\rangle_{\rm conn.} \nonumber\\
&-\frac{1}{2}\int^{t}_{-\infty}d\tau\int^{\tau}_{-\infty}d\tau'\,\Big(Q^{+\,a}(\tau)-Q^{-\,a}(\tau)\Big)\Big(Q^{+\,b}(\tau')-Q^{-\,b}(\tau')\Big)\bigg\langle\{\Phi_{a}(\tau),\Phi_{b}(\tau')\}\bigg\rangle_{\rm conn.}\Bigg]+\cdots\,.
\end{align}

For the purpose of computing decoherence, i.e. the suppression of off-diagonal terms in the reduced density matrix $\bar{\rho}_{q}$, we will only need the magnitude of the influence functional, $\mathcal{F}[Q^{+},Q^{-}]=\exp(-\frac{1}{2}\mathbb{D}[Q^{+},Q^{-}])$. We will henceforth focus on computing the decoherence functional
\begin{equation}\label{eq:generalDecoherence}
\mathbb{D}[Q^{+},Q^{-}]=\frac{1}{2}\int^{t}_{-\infty}d\tau\int^{t}_{-\infty}d\tau'\,\Big(Q^{+\,a}(\tau)-Q^{-\,a}(\tau)\Big)\Big(Q^{+\,b}(\tau')-Q^{-\,b}(\tau')\Big)\bigg\langle\{\Phi_{a}(\tau),\Phi_{b}(\tau')\}\bigg\rangle_{\rm conn.}\,.
\end{equation}

\end{widetext}
\end{document}